\documentclass[sigconf]{acmart}

\newcommand{\sysname}{DesignChecker}
\newcommand{\ipstart}[1]{\vspace{1mm}\noindent{\textbf{\textit{#1.}}}}
\newcommand\smallverb[1]{\texttt{\small #1}}

\usepackage{listings}
\usepackage{makecell}
\usepackage{cprotect}
\usepackage{color}

\newcommand\revised[1]{\textcolor{black}{#1}}

\AtBeginDocument{%
  \providecommand\BibTeX{{%
    \normalfont B\kern-0.5em{\scshape i\kern-0.25em b}\kern-0.8em\TeX}}}

\setcopyright{acmcopyright}
\copyrightyear{2024}
\acmYear{2024}
\setcopyright{rightsretained}
\acmConference[UIST '24]{The 37th Annual ACM Symposium on User Interface Software and Technology}{October 13--16, 2024}{Pittsburgh, PA, USA}
\acmBooktitle{The 37th Annual ACM Symposium on User Interface Software and Technology (UIST '24), October 13--16, 2024, Pittsburgh, PA, USA}
\acmDOI{10.1145/3654777.3676369}
\acmISBN{979-8-4007-0628-8/24/10}

\input{ap.sty}
\begin{document}


\title{DesignChecker: Visual Design Support for \\Blind and Low Vision Web Developers}




\author{Mina Huh}
\affiliation{
  \institution{Department of Computer Science \\ The University of Texas at Austin}
  \country{}}
\email{minahuh@cs.utexas.edu}

\author{Amy Pavel}
\affiliation{
  \institution{Department of Computer Science \\ The University of Texas at Austin}
  \country{}}
\email{apavel@cs.utexas.edu}

\begin{abstract}
Blind and low vision (BLV) developers create websites to share knowledge and showcase their work. A well-designed website can engage audiences and deliver information effectively, yet it remains challenging for BLV developers to review their web designs. We conducted interviews with BLV developers (N=9) and analyzed 20 websites created by BLV developers. BLV developers created highly accessible websites but wanted to assess the usability of their websites for sighted users and follow the design standards of other websites. They also encountered challenges using screen readers to identify illegible text, misaligned elements, and inharmonious colors.
We present DesignChecker, a browser extension that helps BLV developers improve their web designs. With DesignChecker, users can assess their current design by comparing it to visual design guidelines, a reference website of their choice, or a set of similar websites. DesignChecker also identifies the specific HTML elements that violate design guidelines and suggests CSS changes for improvements. 
Our user study participants (N=8) recognized more visual design errors than using their typical workflow and expressed enthusiasm about using DesignChecker in the future.


\end{abstract}



\begin{teaserfigure}
  \centering\label{fig:teaser-2}
  \includegraphics[width=\textwidth]{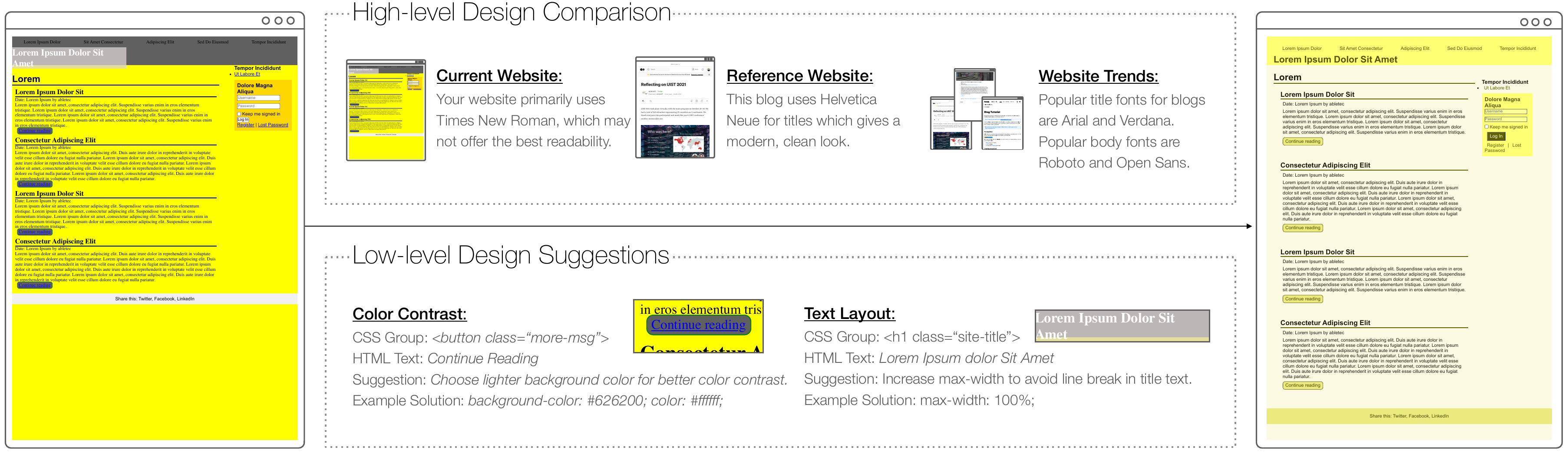}
  \caption{\sysname{} is a browser extension that helps BLV web developers to improve their web design.
  With ~\textit{Design Comparison}, users can easily assess their current web design by comparing it to visual design guidelines, a reference website of their choice, and a trend of similar websites. With ~\textit{Design Suggestions}, users can quickly identify HTML elements with the issue and resolve the issue by following the CSS suggestions.}
\end{teaserfigure}

\maketitle

\section{Introduction}
People create websites to showcase their work, express personal viewpoints, and promote organizations.
Strong visual design attracts user attention and delivers information efficiently~\cite{chen2021umitation, potluri2019ai}. For instance, colors impact first impressions and convey visual hierarchy~\cite{flatla2013sprweb, kikuchi2023generative}, and spacing impacts readability~\cite{lee2020guicomp, riegler2018measuring}. While many BLV web developers create websites to communicate to a broad audience, web design requires visual inspection and thus remains inaccessible.

Recent research enabled BLV web developers to explore the structure of websites using tactile printouts~\cite{li2019editing}, dynamic grids~\cite{li2022tangiblegrid}, and touchscreen displays~\cite{potluri2019multi}.
While such work lets BLV developers access the high-level webpage structure, it remains challenging for BLV developers to assess the visual impact of their design choices such as colors, fonts, and spacing which govern how sighted audiences perceive a website. 
Prior work supported sighted developers to learn from design references~\cite{kumar2013webzeitgeist, kumar2011bricolage, chen2021umitation}, predict user experiences~\cite{reinecke2016enabling, reinecke2013predicting}, and generate and compare design variations~\cite{o2014learning, kikuchi2023generative}.
However, it is not yet explored how to make the design feedback accessible and useful for BLV web developers.



To understand how BLV developers create and review their web design, we interviewed 9 BLV web developers.
While developers create websites to reach a broad audience, their current tools (\textit{e.g.,} screen reader add-ons) did not provide access to many potential design issues. Developers thus recruited sighted reviewers for design feedback, but it was challenging to understand and address their high-level feedback.
To bridge this gap, we first sought to identify the common design challenges BLV developers face. We analyzed 20 websites and found design issues across text (legibility, readability), layout (spacing, alignment), and color (contrast, harmony). For example, sites featured highly saturated yellow text on a blue background (harmony) or center-aligned text for a blog post (alignment). While such issues may be visually evident to a sighted reviewer, BLV developers wanted tools that identify these design issues and suggest how to resolve them.


We present ~\sysname{} (Figure 1), a browser extension that helps BLV developers review and improve the visual design of their websites. ~\sysname{} provides feedback on the text, layout, and color of a web design, as informed by our design analysis of BLV websites. 
With ~\textit{Design Comparison}, users can assess whether their website design follows visual design guidelines via guideline-based feedback, and compare their design with a reference website of their choice or a set of similar websites. 
With ~\textit{Design Suggestions}, users can easily identify the HTML elements related to each design comparison issue and improve the design by following specific CSS suggestions. 
Our pipeline uses algorithms that apply the design heuristics from the guidelines. It processes the website's HTML, CSS, and OCR results to identify issues (\textit{e.g.,} elements with low color contrast) and generate suggestions using GPT-4 (\textit{e.g.,} alternative color values and explanations).

We evaluated ~\sysname{} in a within-subjects study with 8 BLV web developers who compared ~\sysname{} with their current approaches. 
Participants using their own tools made 4.3 (SD=2.1) edits and resolved 2.8 (SD=1.8) design issues while \sysname{} users made 7.8 (SD=2.7) edits and resolved 6.6 (SD=2.8) design issues. Participants using \sysname{} also reported lower frustration and more confidence in their output. All participants expressed they wanted to use \sysname{} in the future as it helped them to quickly identify design issues and make edits. 

This paper contributes the following:
\begin{itemize}
    \item Design opportunities for making visual web design more accessible, derived from a formative study
    \item \sysname{}, a system that supports accessible web design via design comparison and suggestions.
    \item A user study demonstrating how BLV developers use \sysname{} to review and improve their web design.
\end{itemize}



\section{Related Work}
We build on prior work in accessible authoring and programming, as well as research on design feedback and prototyping tools.

\subsection{Accessible Authoring}
Accessibile authoring promotes self-expression for people with disabilities. With the growing interest in accessibility in digital content creation~\cite{zhang2023understanding}, prior work has explored accessible photography~\cite{adams2016blind, adams2013qualitative, jayant2011supporting}, video authoring~\cite{huh2023avscript, li2022exploration}, image generation~\cite{huh2023genassist}, artboard authoring~\cite{zhang2023a11yboard}, and music creation~\cite{payne2020blind}. Studies have also investigated mixed-ability collaboration in slide authoring~\cite{peng2022diffscriber}, document editing~\cite{lee2022collabally}, and programming~\cite{huh2022duoethnographic, pandey2021understanding, potluri2022codewalk, seo2023coding}. 

To enable non-visual authoring, researchers adopted alternative input modalities for authoring tools such as text~\cite{taylor2016customizable, huh2023genassist, huh2023avscript} or audio~\cite{iwamura2020visphoto, zhang2023a11yboard, pandey2020explore}. Other works supported creators to review the content they are authoring with touch \cite{li2019editing, zhang2023a11yboard}, or described visual issues in the images or videos being created~\cite{huh2023avscript, huh2023genassist}.
Building upon this research, we extend the research on accessible creativity support by exploring the realm of visual design for BLV web developers.

\subsection{Accessible Programming}
Studies have explored accessibility in programming including development environments~\cite{potluri2018codetalk}, code navigation~\cite{baker2015structjumper, albusays2017interviews}, and debugging~\cite{albusays2016eliciting}. 
In web development, researchers identified challenges in learning CSS~\cite{kearney2021accessible, pandey2022accessibility}, displaying images~\cite{kearneyaccessible}, verifying colors~\cite{kearney2021making}, utilizing UI frameworks~\cite{pandey2022accessibility}, and understanding layout~\cite{li2019editing, norman2013accessible}. 
To enable access to the layout of webpages, Li et al. used printed tactile printed sheets ~\cite{li2019editing}, and TangibleGrid supported dynamic layouts with an adjustable grid~\cite{li2022tangiblegrid}. To eliminate the need for physical tactile print-outs or grids, Potluri et al. proposed a touchscreen web display~\cite{potluri2019multi} and Borka developed a screen reader add-on for accessing the size and location of UI elements~\cite{NVDAtoolkit}.

While these approaches can capture the spatial layout, they do not provide feedback on the visual impact (\textit{e.g.,} misalignment, visual clutter). Also, it remains challenging for BLV developers to assess additional aspects of web design crucial for user experience. For instance, color contrast impacts text readability~\cite{w3c2023} and font choice impacts the perception of a site's credibility~\cite{selejan2016credibility}, yet these types of feedback are not available to BLV developers. 
In this paper, we tackle the accessibility challenges of web design by supporting BLV developers to identify and address common design issues.

\subsection{Design Feedback}
Feedback helps novices understand key principles in a domain, articulate their goals, and recognize how others will perceive their work~\cite{feldman1994practical}. 
Prior work leveraged crowdsourcing to generate and visualize design critique~\cite{luther2015structuring, xu2014voyant, yuan2016almost}. 
To enable designers to foresee users' responses to a design, researchers predicted emotional responses to different font styles in web pages~\cite{bianchi2021emotional} and quantified perceived visual complexity and colorfulness~\cite{reinecke2013predicting}. ColorCheck, an image-processing tool helped designers to understand which colors users cannot see under different situations~\cite{reinecke2016enabling}.

Recent works investigated the evaluation of graphic design based on design principles and user goals~\cite{o2014learning} and supported users to compare their design with a dataset of examples with visualization of distribution~\cite{lee2020guicomp}. 
While most design feedback tools use visual markers and demonstrations (\textit{e.g.,} drawing a box around the relevant element)~\cite{duan2024generating, kim2022stylette, xu2014voyant}, design feedback for BLV developers requires different considerations.
To our knowledge, no prior work explores how to effectively communicate visual design issues and solutions to BLV developers. Our work builds upon the rich prior work on design feedback, exploring accessible visual design feedback.

\subsection{Web Design Prototyping Tools}
To support end-users to easily modify their visual web design, various tools have been developed. AI-powered tools enabled novice designers to explore the alternative styles of the web~\cite{kim2022stylette, shi2023stijl}. Example-based design tools were introduced to facilitate the search for design inspirations~\cite{lee2020guicomp, swearngin2018rewire} and reduce the mental and manual effort of copying other websites' styles~\cite{chang2012webcrystal, chen2021umitation, kumar2013webzeitgeist, kumar2011bricolage}.
To support the design of an accessible artifact, SUPPLE explored the automatic generation of user interfaces that adapt to people with motor impairment~\cite {gajos2004supple} and SPRWeb created accessible web design for people with color vision deficiency via automatic recoloring~\cite{flatla2013sprweb}. 
Existing web accessibility evaluation tools and vision simulation tools help developers determine if their web content meets accessibility guidelines~\cite{w3c_wai_2023}.
Drawing inspiration from such tools, we explore how a design evaluation tool can help BLV developers create websites that better adhere to the visual design guidelines.

\section{Formative Work}
To understand current practices and challenges for web design, we first conducted a semi-structured interview with BLV developers. 
We additionally conducted a design analysis of websites created by BLV developers to understand the common design issues.

\subsection{Interview}

We recruited 9 BLV creators experienced in creating websites (P1-P9, Table~\ref{tab:participants}). Participants were recruited using mailing lists and compensated 20 USD for the 30-minute remote interview conducted via Zoom\footnote{Approved by the institution's Institutional Review Board (IRB).}. 
We asked participants how they currently decide on, implement, and review their website designs, and what accessibility barriers they encountered with their current approaches. 

\ipstart{Participants} 
Participants were totally blind (5 participants), or legally blind (4 participants). All of them used screen readers (NVDA, JAWS), and 3 of the participants occasionally used magnifiers with their remaining vision (P1, P7-P8). 7 participants created websites for clients as part of their profession (P2-P3, P5-P8), 6 participants created websites to showcase their personal work (P1, P3-P4, P7-P9), and 2 participants volunteered to build websites for non-profit blind organizations (P2, P4). 
All participants designed websites for both BLV and sighted visitors, and 4 participants also had experience creating websites designed primarily for BLV visitors (P2, P4-P5, P8). 
Participants designed their websites by coding their websites from scratch (P1-P3, P5-P6, P8-P9), adapting existing templates (P1-P7, P9), and using static site generators (P5, P9) or block-based editors (P7). All participants created websites using HTML and CSS, and 6 participants used JavaScript for interactivity (P2-P6, P8). 
Only one participant mentioned using a UI framework (Bootstrap) offering native UI components (P7).

\ipstart{Motivations for Web Design}
Prior work revealed the motivation of BLV developers for general UI development (\textit{e.g.,} hiring opportunities, effective communication with designers~\cite{pandey2022accessibility}). Our work further explores why visual design is important for BLV developers.
All participants wanted to create websites that are accessible and visually appealing for both BLV and sighted people. 
Participants mentioned 4 motivations for creating visually appealing websites.
First, they wanted to make their websites more interesting to sighted people (P1-P3, P8-P9). P2 said ~\textit{``So that sighted people would enjoy my website too! My daughter, my wife, I want anyone to find it interesting!''} Second, participants wanted to provide a better user experience for sighted people (P6-P7, P9). P7 noted they wanted sighted users to have ~\textit{``a smooth experience''}, explaining ~\textit{``If a website doesn't look pretty, you get to focus on the rough edges and less on the content or task.''}
Similarly, P9 mentioned \textit{
``If I am asking sighted people to make their websites accessible, I should do it for them too!''}. Third, participants wanted to gain confidence in sharing with sighted audiences (P4-P5). 
Finally, they wanted to increase viewership (P2, P6). P6 said ~\textit{``Look is equally valuable as the content. If one is missing, no one will visit.''} 



\ipstart{Learning Web Design}
Prior research highlighted the inaccessibility of learning materials for web design (\textit{e.g.,} example images without alt text, video tutorials without audio descriptions)~\cite{kearney2021accessible, pandey2021understanding}. 
\revised{Our participants further noted that it was difficult to use the code preview tools common in online tutorials (\textit{e.g.,} CodePen, CodeSandbox) to understand design changes as the editors and preview panels of these tools are not screen reader accessible.} While some tutorials featured text descriptions of visual design concepts, they did not describe specific visual effects of such concepts to support BLV developers. P8 stated ~\textit{``I read many many documents, but still it is so hard to know when to use borders, picture frames, or margins!''}

\ipstart{Searching for Web Design Inspiration}
Participants searched online to get inspiration from existing websites with a similar purpose (P2, P5-P8). Yet, it was time-consuming to explore the designs by code (P5-P6) and difficult to select a reference without understanding how each design looks (P6, P8). 
P8 described, ~\textit{``I can't see others' websites. I don't see the decorations or anything related to style with a screen reader!''} 
Echoing findings from Li et al.~\cite{li2019editing}, participants also searched for web templates (P1-2, P4-P5, P7, P9) for quick development. Yet, templates were often not screen reader accessible (P1, P5-P6) or lacked descriptions of how they look (P1, P4-P7).
\revised{Thus, 5 participants selected templates based on only their accessibility and functionality without considering how the template design looked (P1-2, P4-6).
P5 noted ~\textit{``I just randomly picked one assuming the sighted person who created it already did well''}, but P1 and P4 felt less confident in the look of their template and in sharing their final websites with sighted people. P2 asked a sighted person to describe the look of his website after modifying the template.
P9 uniquely considered template appearance first by asking a sighted person to find a template with specific features and a visual theme (\textit{e.g.,} clean and elegant), then asking the sighted person to make the template code accessible by adding \smallverb{aria} labels so that he can fill in the content. 
4 participants modified the content or style of the template after selecting one (\textit{e.g.}, P1 and P2 removed unnecessary UI elements, P4 and P7 updated colors to match their preferences).}

\ipstart{Reviewing Web Design}
When reviewing their web design, participants wanted to confirm the changes made (P1, P5-P9), identify design issues (P1-P9), and compare their web design to other similar websites (P1-P2, P5). 
In line with earlier findings~\cite{pandey2021understanding, pandey2022accessibility, li2019editing, kearney2021accessible}, BLV developers utilized screen readers or add-ons to check the navigation order (P1, P5-P7, P9) or color of web elements (P1, P5, P8). P5 described ~\textit{``For colors, I use an NVDA add-on [audioScreen]}~\cite{audioScreen.md} ~\textit{that converts the RGB values to sound. [...] While it can help me verify whether a change is applied, it is not helpful for making decisions.''}

3 participants also used their remaining vision to review the web design by magnifying the screen (P1, P7-P8). P7 and P8 mentioned that they adapted their design to make reviewing easier. P8 mentioned ~\textit{``I always make the borders of elements real big and increase the color contrast so that I can clearly distinguish elements and review them by myself.''} Other participants used touch displays to understand the layout (P5), HTML and CSS validators to identify issues in rendering (P6), and an Image-to-Sound AR glass~\cite{seeingWithSound} to understand the rough layout and contrast between elements (P9).
While participants' current strategies improved their understanding of the high-level layout (\textit{e.g.}, screen reader order or touch display) and low-level details of the design (\textit{e.g.}, audioScreen reading color hex values), the strategies did not support recognizing many design issues that a sighted person may notice. 

\ipstart{Understanding Design Feedback}
8 participants asked for sighted people's help in reviewing their web design (P1-P3, P5-P9) as shown in prior work~\cite{pandey2021understanding, li2019editing, norman2013accessible, mealin2012exploratory}.
Participants sought feedback from average website users (P1-P2, P7, P9), people with programming backgrounds (P3, P5-P6), or assistance services for visual interpretation such as Aira~\cite{Aira} (P7-P8). 
P4 did not ask sighted people to review his website due to his lack of confidence~\textit{``I'm not confident enough to show my current design to any sighted people. I'll be more comfortable if it's not human that's giving me feedback.}

We also reveal that participants commonly struggled with interpreting high-level feedback. P6 noted ~\textit{``She [sighted reviewer] says ``This doesn't look good now.'' I always have to ask for clarifications and elaboration.''} 
Similarly, P3 found it challenging to come up with what CSS changes to make to address the feedback.
P1 mentioned that sighted reviewers are not always forthcoming with their reviews. ~\textit{``[Sighted] people who are reviewing don't want to be too negative when they give feedback. I know my website doesn't look nice, but they try to be nice.''}
Participants also raised the concern of not being able to validate the reviews themselves. P1 stated ~\textit{``Even if their feedback is not correct, or even a bad idea, I don't know their artistic level and cannot interpret it.''}

\ipstart{Comparing to Web Design Standards} 
In addition to identifying specific design issues, participants wanted to assess if their website followed the visual design standards of other websites. 
P2 described ~\textit{``Standards are important. If mine doesn't look like other websites, users will not easily find what to do and won’t visit my website again.''} 
Specifically, participants were interested in websites with similar contexts and purposes. 
P1 said ~\textit{``If it's a business website, I want to compare it with websites that have a professional look. Not something with many decorative images.''} 
In describing an ideal system to support web design, P5 noted ~\textit{``I want to give some sample blog websites and tell it to make mine look similar. But I don't want it to exactly copy their styles because I want to control what to follow.''}

\ipstart{\revised{Implementing Web Design}}
\revised{Among the 8 participants who coded their websites from scratch, 7 of them first developed features and content then added style. Only P3 planned the layout early using his web development knowledge from before losing sight. 
To design and implement their websites, P5 and P6 asked collaborators (\textit{e.g.,} UI designers) or clients to specify the design in detail (\textit{e.g.,} gray button with rounded corner) then referred to tutorials (\textit{e.g.,} w3school) to implement it. P3 and P8 instead started with a design in mind (\textit{e.g.}, layout, colors) and asked sighted people how to implement it, but they reported that sighted people's instructions were difficult to follow. 
P9 who coded independently searched for sample stylesheets with descriptive class names (\textit{e.g.,} menu-item), copied the code, and matched the names to his UI components.}

Participants also edited their web design to follow styles from other websites (P2, P6) and address design feedback that they received (P1-P3, P5-P9).
Yet, P6 expressed hesitation about re-using other websites' design code ~\textit{``If I add my content instead of the original text, the length of the element changes and might look weird.''}
When making changes based on the feedback, participants wanted to confirm whether the changes addressed the feedback and did not introduce additional design issues, however immediate feedback after each update was lacking (P2, P5-P6, P9).



\subsection{Design Opportunities}\label{sec:design-opportunities}
From the formative interview, we derive the following design opportunities to make web design more accessible to BLV web developers:

\begin{itemize}
    \item[\textbf{D1.}] Provide accessible resources for learning visual design.
    \item[\textbf{D2.}] Support users to search for design inspirations and templates.
    

    \item[\textbf{D3.}] Support users to assess a web design in comparison to other websites with similar contexts and objectives.

        \item[\textbf{D4.}] Identify visual design issues and describe their impact.
    
    \item[\textbf{D5.}] Pinpoint elements with issues and provide code-level improvement suggestions.
    
    \item[\textbf{D6.}] Describe the changes made and their impact on the design.
\end{itemize}

The design opportunities address challenges in learning web design (\textbf{D1}), searching for references (\textbf{D2}), reviewing (\textbf{D3}, \textbf{D4}), and updating (\textbf{D5}, \textbf{D6}) the web design. In this work, we explore how to make web designs more accessible by supporting BLV web developers to assess their visual design (\textbf{D3}, \textbf{D4}) and make changes for improvements (\textbf{D5}, \textbf{D6}), while the remaining challenges (\textbf{D1}, \textbf{D2}) can catalyze future research. 
While some design opportunities may seem universally applicable to web development, they are particularly crucial and nuanced for BLV developers and require a fundamentally different approach.
In this work, we explore design issues common in BLV developers' websites and provide non-visual design feedback on these issues that are navigable and interpretable through assistive technologies.

\subsection{Website Design Analysis}~\label{sec:website-analysis}
Our interview participants wanted support to identify visual design issues in their websites (\textbf{D4}). To identify the common design issues, we conducted a design analysis of websites created by BLV developers. We collected the websites through publicly archived mailing list posts from the discussion group of BLV computer programmers~\footnote{https://www.freelists.org/list/program-l (Jul.2021-Jul.2023)}. 
For websites with multiple pages, we only added pages that are substantially different in visual layout (\textit{e.g.,} landing page, directory page, post page). The final dataset had 20 websites including 10 blogs/tutorials, 4 organization websites, 2 podcast websites, 2 product websites, and 2 personal websites. 

Two researchers experienced in web design critique conducted the analysis based on the design principles of Luther et al.~\cite{luther2014crowdcrit, luther2015structuring} (details in Appendix~\ref{apndx:website_analysis}). We reveal common design issues and related CSS properties: legibility (\smallverb{font-size}), readability (\smallverb{font-family, line-spacing}), spacing (\smallverb{margin, padding}), alignment (\smallverb{text-align, margin, padding}), color contrast (\smallverb{background-color, color}), and color harmony (\smallverb{background-color, color, border-color}).
Table~\ref{tab:form-analysis} shows the number of identified design issues. For assessing color harmony, we did not count the number of elements with the issue but instead rated holistically (as the issue is determined by a website as a whole rather than from individual elements). 


\begin{table}[t]
\sffamily\def\arraystretch{0.7}\setlength{\tabcolsep}{0.4em}
\centering
\begin{tabular}{lcccccc}
\toprule
 & \multicolumn{2}{c}{Text (\#)} & \multicolumn{2}{c}{Layout (\#)} & Color (\#) \\
\multicolumn{1}{c}{} & \multicolumn{1}{l}{Legibility} & \multicolumn{1}{l}{Readability}  & \multicolumn{1}{l}{Spacing} &   \multicolumn{1}{l}{Alignment} & Contrast\\ \midrule
 \textbf{$\mu$} &  4.05  & 2.55 & 3.15 & 4.00  &  2.40 \\
 \textbf{$\sigma$} &  3.39 & 1.57 & 2.98 & 3.52 &  1.64\\
 \bottomrule
\end{tabular}
\caption{We report the number of text, layout, and color issues identified in 20 websites created by BLV developers. For color harmony, we rate holistically instead of counting issues (2 had good, 11 had moderate, and 7 had weak harmony).}
    \label{tab:form-analysis}
    \vspace{-5pt}
\end{table}

\section{\sysname{}}

\begin{figure}[t]
  \centering
  \includegraphics[width=0.7\columnwidth]{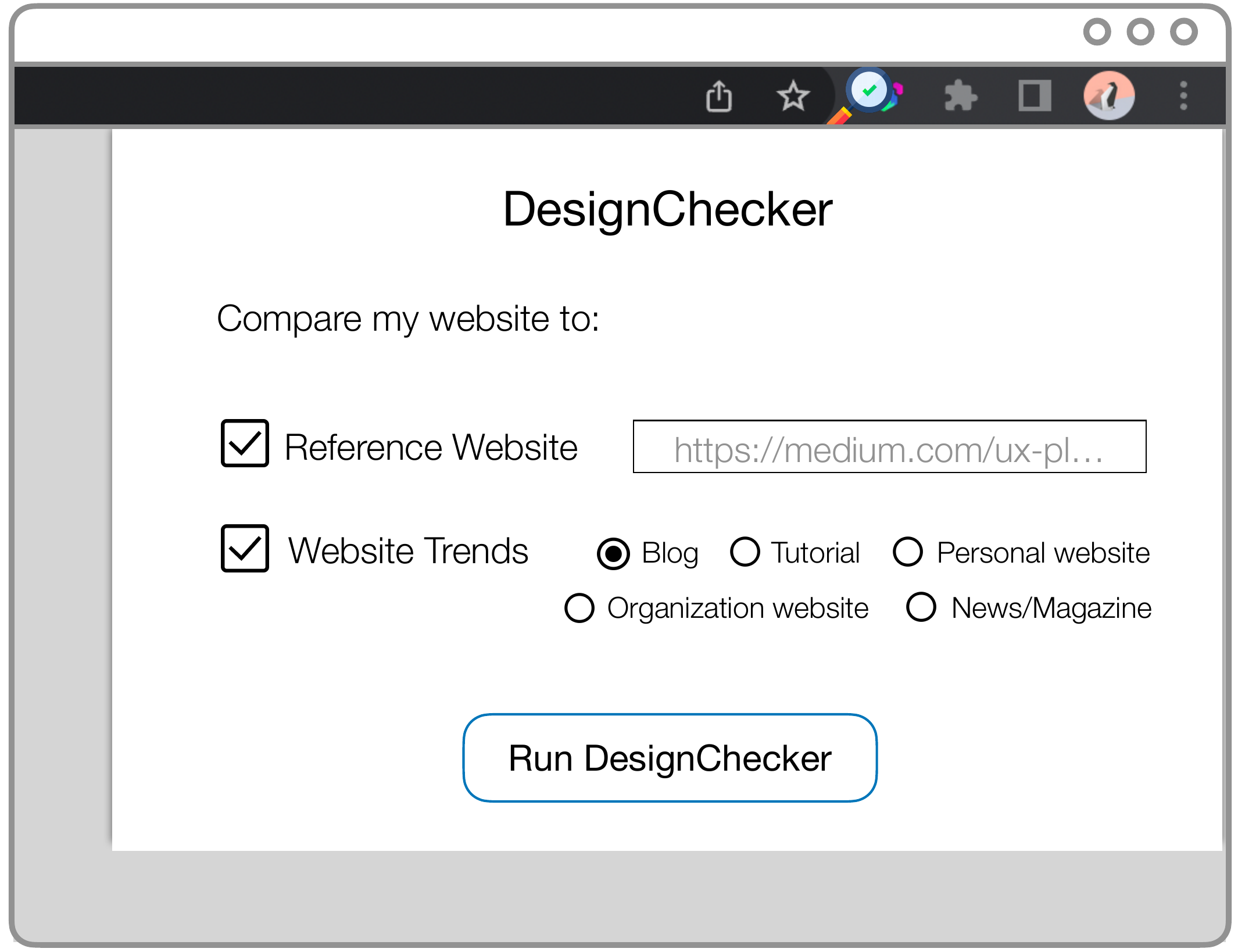}
  \caption{When activated, ~\sysname{} overlays a popup where users can select the options for ~\textit{Design Comparison}:  a reference website and website trends.}
  \label{fig:system_popup}
\end{figure}

\begin{figure*}[t]
  \centering
  \includegraphics[width=0.85\textwidth]{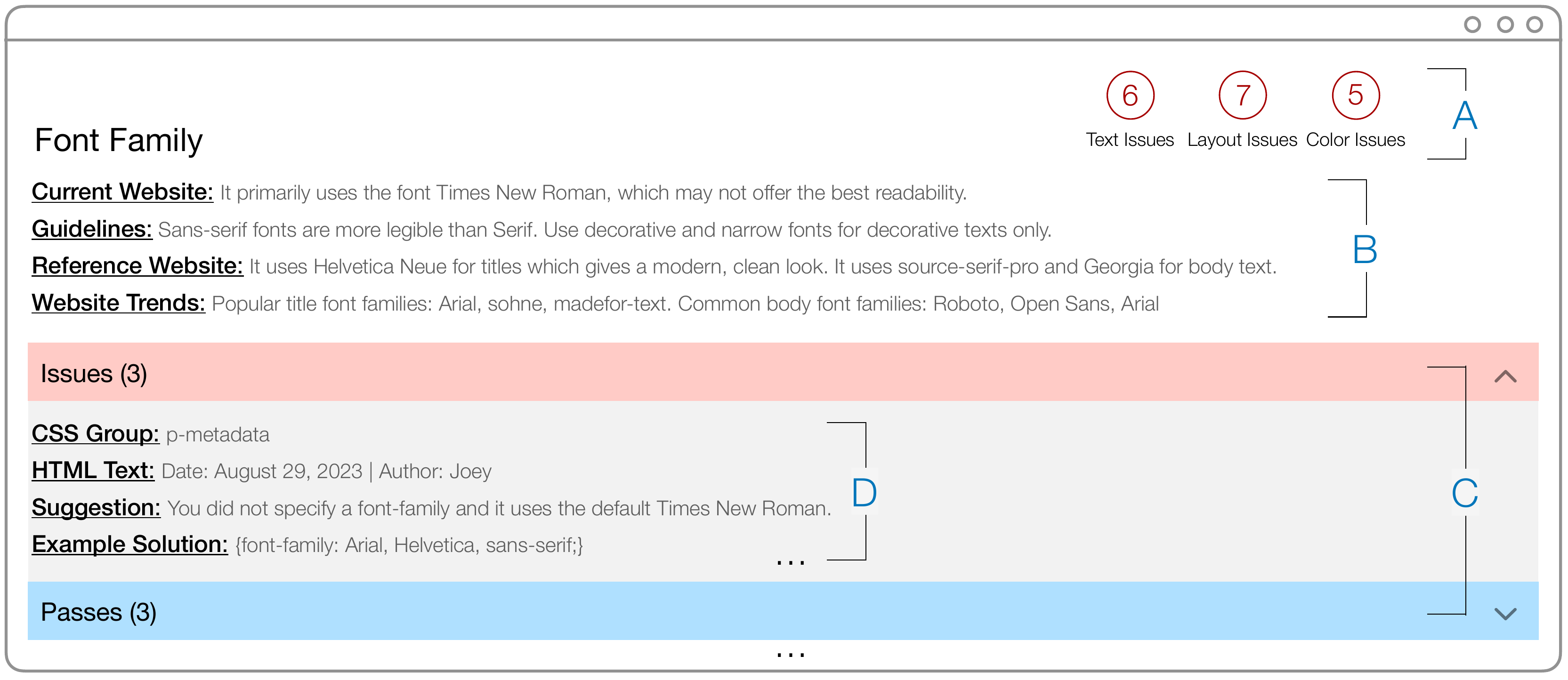}
  \caption{\sysname{} is a browser extension that helps BLV developers to review and improve their web design. ~\textit{Design Issue Summary} (A) provides the overview of issues in the current design. ~\textit{Design Comparison} (B) enables users to assess their current design using 3 different resources: visual web design guidelines (default), a reference website of their choice, and a trend of similar websites. ~\textit{Design Suggestions} (C) provides the list of elements that failed (Issues) or fulfilled (Passes) the requirements of design guidelines. Each ~\textit{Design issue} (D) identifies the specific HTML elements causing the issue and suggests CSS changes.}
  \label{fig:system_overview}
\end{figure*}

We present \sysname{}, a browser extension that helps BLV web developers improve their web design by comparing it to references, identifying design issues, and providing suggestions (\autoref{fig:system_overview}). 
To address \textbf{D3}, \sysname{} supports high-level ~\textit{Design Comparison} that enables a high-level comparison of the current web design with a reference website of users' choice and a trend of similar websites (Figure \ref{fig:system_overview}B). 
To address \textbf{D4} and \textbf{D5}, \sysname{} provides low-level ~\textit{Design Suggestions} to enable users to easily identify the specific elements related to each issue and follow suggestions (Figure \ref{fig:system_overview}C). 
The types of design issues that ~\sysname{} addresses are informed by the web design analysis (Section~\ref{sec:website-analysis}) -- evaluating common design issues found in BLV web designs including text (legibility, readability), layout (whitespace, spatial alignment, textual alignment), and colors (color harmony, color contrast).
To address ~\textbf{D6}, ~\sysname{} provides the summary of code changes and the resulting design changes.

\subsection{Interface}
To illustrate how BLV web developers can use ~\sysname{}, we follow Joey, a web developer who uses a screen reader to create websites. While she has created multiple websites before, she never felt confident about sharing her websites with sighted friends because of the websites' visual design. Recently, she posted several blog posts that will not only help BLV web developers but also sighted developers, and wants to promote her blog to a broad audience. To improve her web design, she first looks up web design guidelines online but soon realizes that most suggestions are high-level (\textit{e.g., ``Make sure your website design is consistent.'', ``Pay close attention to the balance of elements.''}) and thus difficult to directly apply. She decides to use ~\sysname{} to re-design her blog website to deliver the content more effectively and appear more professional.

After installing ~\sysname{}{}, she opens her blog website and clicks on the extension icon to open the popup (Figure~\ref{fig:system_popup}).
To get rich feedback, she selects both design comparison source options: a reference website, and a trend of multiple websites. For the reference website, she enters the link to a blog post from Medium UX Planet because it has many followers and covers similar topics to her blog. For the website dataset type, she selects `Blog' as she is currently interested in what other blog websites' styles look like rather than all other types of websites. 
Once she clicks the ~\textit{`Run DesignChecker'}, ~\sysname{} runs a design analysis of the current page and opens the results in a new window.

\ipstart{Design Issue Summary}
The design issue summary provides the number of issues in each design category (Figure~\ref{fig:system_overview}A). 
Joey realizes that her current web design has design issues in all text, layout, and color categories, which were not identifiable with her screen reader.
To resolve each of the design issues, she reviews the design comparison and design suggestions.

\ipstart{Design Comparison}
With \sysname{}'s design comparison (Figure ~\ref{fig:system_overview}B), users can easily assess their current web design using 2 different resources: a reference website of their choice, and a dataset of similar websites. For each design category (\textit{e.g.,} readability), \sysname{} provides a summary of users' current website (\textit{''It primarily uses the font Times New Roman, which may not offer the best readability.''}), and the guideline recommendations (\textit{``Sans-serif fonts are more legible than Serif. Use decorative and narrow fonts for decorative texts only.''}).
For users seeking comprehensive details, links to the full guidelines (Table ~\ref{tab:form-analysis}) are provided. 
Based on the options users selected in the popup interface, ~\sysname{} provides a summary of the reference website and a summary of multiple websites.
Joey learns that not specifying a font family caused most text elements to use the default `Times New Roman', which may not provide ideal readability for web content. She checks the blog dataset comparison to see the trends of font families used in other blog websites. Then, she decides to use `Open Sans', the most popular font family used for body text. 

\ipstart{Design Suggestions}
\sysname{}'s design suggestions provide the list of elements that failed (\textit{Issues}) or fulfilled (\textit{Passes}) the requirements of design guidelines  (Figure~\ref{fig:system_overview}C). 
Unlike most design feedback tools that use visual markers (\textit{e.g.,} drawing a box around the relevant element)~\cite{duan2024generating, kim2022stylette, xu2014voyant}, \sysname{} focuses on the non-visual presentation of suggestions that is easy to interpret and address. 
\sysname{} presents the issues in the unit of a~\textit{`group'}, which represents a group of HTML elements that share the same style attributes (Figure~\ref{fig:system_overview}D).
The ``CSS group'' information helps users to easily notice which parts of the code to edit, and the ``HTML Text'' information (\textit{e.g.,} inner text or alt text for images) helps users remember the usage of the group.
The grouping is also beneficial for users because it does not repeat the identical suggestions for multiple HTML elements that apply the same CSS styles.
\sysname{} also lets users access elements without issues (\textit{passes}) via a collapsable menu. 
From design suggestions, Joey discovers that 3 text groups are associated with font family issues. 
Because \sysname{} pinpoints which HTML elements and CSS styles are related to each issue, she can easily find all text elements with font family issues and make updates.

\begin{figure}[t]
  \centering
  \includegraphics[width=\columnwidth]{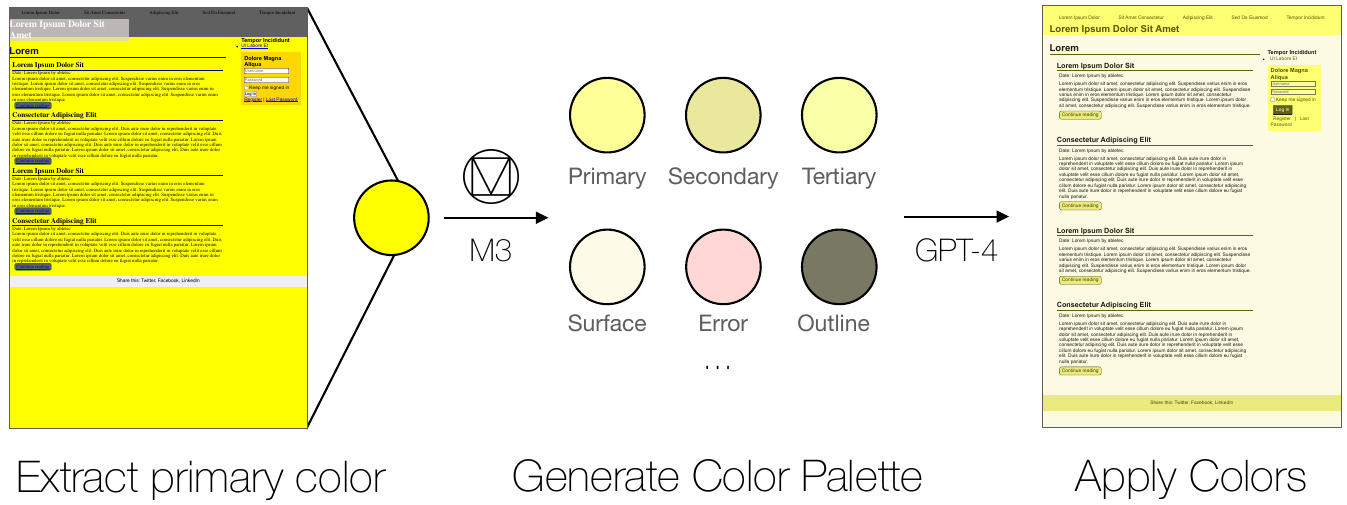}
  \caption{To improve color harmony, \sysname{} first extracts the primary color from the original web design and generates a color palette for pre-defined roles (\textit{e.g.,} surface, outline)~\cite{MaterialsColor}. Using the definitions of color roles and the generated palette, GPT-4 suggests colors for UI elements.}
  \label{fig:system_color_audit}
\end{figure}

To improve color harmony, ~\sysname{} provides a single holistic suggestion that spans multiple elements, because color harmony depends on the collective use of colors across multiple elements (Figure~\ref{fig:system_color_audit}). First, \sysname{} provides the summary of the current color scheme to motivate and explain the edit suggestions. Then, based on the primary color in the original website, it generates a new color palette and describes where to apply new colors.
Joey reviews ~\sysname{}'s suggestions to improve color harmony. When designing the colors of the website, she chose a blue background with yellow text to provide high contrast. However, from the design suggestions, she realizes that a background with high saturation does not look good and that using too many colors can give a cluttered feel. Following the suggestions, she uses lighter blue for the background and changes the font color to black.


After addressing all other suggestions from ~\sysname{}, Joey saves the edited code in the code editor, which automatically triggers ~\sysname{} to run a new analysis on the updated visual design. On top of the new results page, ~\sysname{} provides the summary of code changes and the resulting issue summary (number of issues resolved or introduced with the new design). She notices that all design issues are resolved, and confidently shares the link to her blog on social media.

\subsection{Pipeline}
\begin{figure*}[!htbp]
  \centering
  \includegraphics[width=0.8\textwidth]{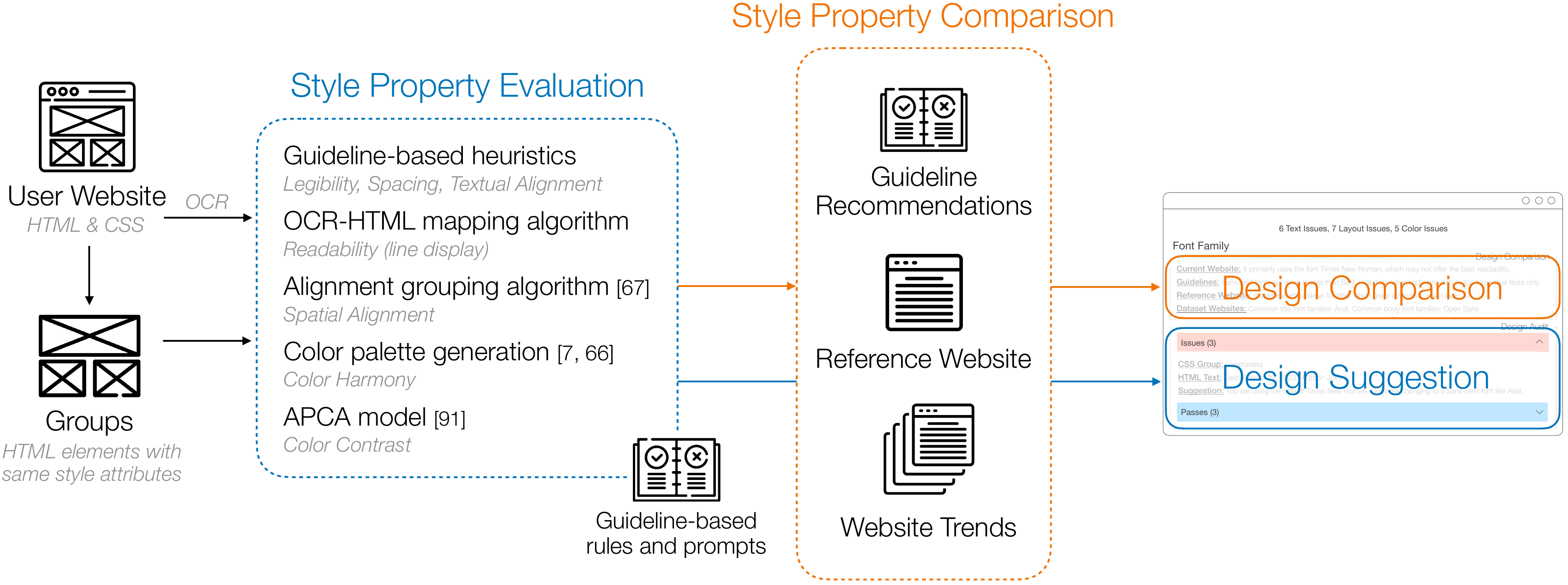}
  \caption{Pipeline of ~\sysname{} that generates ~\textit{Design Comparison} and ~\textit{Design Suggestions} from the user's website. From the input website, ~\sysname{} extracts HTML and CSS and groups HTML elements that share the same CSS classes. 
  Identified issues and the relevant guideline suggestions are presented in ~\textit{Design Suggestion}.
\sysname{} compares the current website's evaluation results with the recommended values in the guidelines, the reference website, and the website trends. The summary of the comparison is presented in ~\textit{Design Comparison}.}
  \label{fig:system_pipeline}
\end{figure*}

To enable Design Comparison and Design Suggestions, our pipeline extracts HTML, CSS, and OCR results from the input website and runs style property evaluation (\autoref{fig:system_pipeline}). 

\subsubsection{Style Property Evaluation}~\label{pipeline_design_audit}

\ipstart{Text: Font Size}
\sysname{} extracts all text elements and compares the font size of the title and body elements to the recommended font sizes (20px and 16px, respectively).

\ipstart{Text: Font Family}
\sysname{} uses GPT-4~\cite{openai2023gpt4} to review the current font families and give suggestions. We prompted GPT-4 with the readability guideline related to font family (\autoref{tab:form-analysis}) and provided a single shot example to specify the input and output format for automating the pipeline (full prompt in Appendix B.1). We used GPT-4 instead of using a rule-based approach with pre-curated sets of ~\textit{``good''} and ~\textit{``bad''} font families because it is difficult to collect all existing font families and judging their readability based merely on being Serif or Sans Serif can be too restrictive.

\ipstart{Text: Line Length}
\sysname{} checks whether the titles do not have a line break and the line lengths of body text are 50-75 characters long. While text line length affects readability, the code does not reveal how text elements are displayed on the rendered page and the lines they encompass. 
To get effective line information of text, we ran Google's Document AI OCR (Optical Character Recognition)~\cite{google_documentai} on the screen capture of the website. We developed an algorithm to match the text elements to the lines of the OCR response of a page. We consider the following match between an HTML text element and an OCR line: 1:1 (where text spans a single line and is the only element in the line), 1:N (where text is a long paragraph and spans multiple lines), N:1 (where multiple text elements aligned horizontally in a single line). 

Using the OCR's dimensions \( w \) and \( h \) alongside the original canvas dimensions \( w_o \) and \( h_o \), the algorithm computes scaling multipliers: \(\text{scalingFactorX} = \frac{w_o}{w}\) and \(\text{scalingFactorY} = \frac{h_o}{h}\). These adjust normalized OCR vertices to pixel positions on the primary canvas. Then, we calculate the matching score between each text element and OCR line using proximity and text similarity. For proximity, we calculate the intersection area of bounding boxes or distance with a threshold of 50px. For text similarity, we use a modified Levenshtein Distance~\cite{yujian2007normalized} to evaluate word correspondences. While the original approach penalizes differences in text length, we adapted the algorithm to allow 1:N or N:1 matching by evaluating similarity scores using length ratios.

\ipstart{Layout: Margin \& Padding}
~\sysname{} computes the margin and padding of HTML elements that are meant to be~\textit{visible} --- ignoring metadata tags (\textit{e.g.,} \smallverb{script}, \smallverb{style}, \smallverb{link}), and zero opacity or zero dimension elements. Following the guidelines on using margin and padding for visual web designs~\cite{uxengineer2023, materialio2023}, we checked whether the container elements (\smallverb{div}, \smallverb{section}, \smallverb{article}, \smallverb{nav}, \smallverb{aside}, \smallverb{header}, \smallverb{footer}, \smallverb{main}, \smallverb{button}, \smallverb{input}, \smallverb{textarea}) have minimum recommended padding of 24px, and atomic elements (\smallverb{p}, \smallverb{h1}, \smallverb{h2}, \smallverb{h3}, \smallverb{h4}, \smallverb{h5}, \smallverb{h6}, \smallverb{li}, \smallverb{a}, \smallverb{button}, \smallverb{input}, \smallverb{label}, \smallverb{img}) have minimum recommended margin of 8px between each others. 
Because combining margin and padding together can break the alignment on top and left, we follow ~\textit{down-right} method~\cite{uxengineer2023} to use \smallverb{margin-bottom} or \smallverb{margin-right}.

\ipstart{Layout: Alignment}
~\sysname{} checks the textual alignment of text elements and suggests left-alignment for bullet texts (\textit{e.g.,} \smallverb{li}) and paragraph texts for readability~\cite{felterunfiltered2023}. 
To check the spatial alignment between close UI elements, we follow Donovan et al.~\cite{o2014learning}'s algorithm to find loose alignment groups of 6 possible types: Left, X-Center, Right, Top, Y-Center, and Bottom. To form alignment groups, we first compare a pair of HTML elements and their bounding boxes and check whether the difference in edge (Left, Right, Top, Bottom) or center (X-Center, Y-Center) is below a threshold of 5. Text blocks can only align with elements of the same alignment type (e.g., left-aligned text can't join an X-Center group). An alignment group is a connected set of aligned items, but elements separated by another aren't grouped together. After the groups are identified, \sysname{} finds slightly misaligned elements that have different edge or center than other elements in the group and suggests changing their position or margin to match the alignment.

\ipstart{Color: Harmony}
\sysname{} provides a summary of the current color scheme by first extracting colors from elements with a color set in CSS (\textit{e.g.} \smallverb{color}, \smallverb{background-color}, \smallverb{border-color}) then extracting 3 primary colors from images using Color Thief~\cite{color-thief}. As BLV developers often can not preview colors and prefer to read color names over raw rgb or hex values~\cite{kearney2021accessible, kearney2019blind}, we translate our extracted color values into interpretable color names (\smallverb{rgb(168, 180, 255)} to \textit{Soft pastel blue}). 
Existing online tools for color mapping offer only a subset of color codes~\cite{njitColor}, unintuitive naming conventions such as numbering (\textit{e.g., grey1-grey99}~\cite{njitColor}) or abstract names (\textit{e.g., island dream}~\cite{colornamer}), thus we use GPT-4 for color translation as prior work~\cite{colorGPT} (full prompt in Appendix~\ref{apndx:color_scheme_prompt}).

To provide color harmony suggestions, ~\sysname{} considers BLV developers' original design choices by re-using the primary color used in the original design and also matching the light/dark mode. 
To extract the primary color, \sysname{} extracts 5 dominant colors from the screenshot~\cite{color-thief} and sets the first non-achromatic colors (\textit{e.g.,} non-white or black) as the primary color. 
With the primary color, the pipeline uses Material Design's color theme generator~\cite{MaterialsColor} to generate palettes for multiple roles (\textit{e.g.,} primary, secondary, surface, on surface, outline, error)~\cite{ColorRoles}. \sysname{} determines the light/dark mode based on the greyscale value of the original background color and generates palettes accordingly (light mode: Figure~\ref{fig:website_results}A, D, dark mode: Figure~\ref{fig:website_results}B-C, E).

\ipstart{Color: Contrast}
Although feedback on color harmony suggests colors that offer sufficient contrast, we provide separate feedback on color contrast because websites can have contrast problems without lacking harmony. Also, users may prefer to address color contrast concerns without altering their overall color scheme, making local adjustments instead.
To determine the color contrast for legibility, ~\sysname{} uses APCA (Accessible Perceptual Contrast Algorithm)~\cite{w3c2023apca} that predicts the contrast for web standards (WCAG 3). Using the APCA model, we calculate the LC (Lightness Contrast) value which decodes the perceptual lightness from text and background color, then factor other attributes including font size, font weight, and the usage (\textit{e.g.,} body, title, decorative text). For text elements with an LC value lower than the recommended threshold of 74.7, we suggest alternative colors for text or background that achieve higher contrast. 

\subsubsection{Style Property Comparison}
To implement Design Comparison, our pipeline first summarizes the design of the current website by providing the most common values for font size, line length, margin \& padding, alignment, and color contrast (\textit{e.g.,} \textit{``Your most common font size for the title is 24px, and body is 13px.''}). For font family and color harmony, we use GPT-4 to generate the summary (detailed in Section~\ref{pipeline_design_audit}).

Then, \sysname{} runs ~\textit{style property evaluation} on the reference website and selected categories of website trends (Blog, Tutorial, Personal Website, Organization Website, and News/Magazine). To generate the trends of 5 categories, we constructed a dataset of 100 websites (See Appendix~\ref{apndx:website_dataset} for details).
For design attributes with numerical value suggestions (\textit{e.g.,} font size, margin, padding), \sysname{} provides the most frequent values with the range. For the line length and color contrast category, we summarize the types of elements of the reference website that violate the guidelines (\textit{e.g., ``The reference website has color contrast lower than the minimum recommended contrast in the input tag.''}). We chose to flag the element types with issues on the reference website rather than presenting their exact values (\textit{e.g.}, average color contrast value or line length) as these values may not be as easily applicable to users' sites.
For font family and color harmony, \sysname{} uses GPT-4 to map the rgb and hex values to color names and to describe the feel of the color schemes and the font families used. 
For example, it describes ~\textit{``The reference website is using white (rgb(255, 255, 255)) for background and dark green (rgb(0, 30, 29)) for text and medium sea green (rgb(64, 186, 155)) for borders.''} For comparison with the website dataset, we summarize based on the most common font families and colors across websites. 


\subsection{Implementation}
We implemented ~\sysname{} as a Chrome extension using JavaScript, HTML, and CSS. It uses Mozilla's Window API to extract the HTML and CSS in real-time and uses Chrome's extension API to capture the rendered screen of websites. For the backend, we used a Node.js server to process requests to Google's Document AI OCR APIs~\cite{google_documentai}. We tested the compatibility of ~\sysname{} with three major screen readers: NVDA, JAWS, and VoiceOver.
\section{Results}
\begin{figure*}[!htbp]
  \centering
  \includegraphics[width=0.9\textwidth]{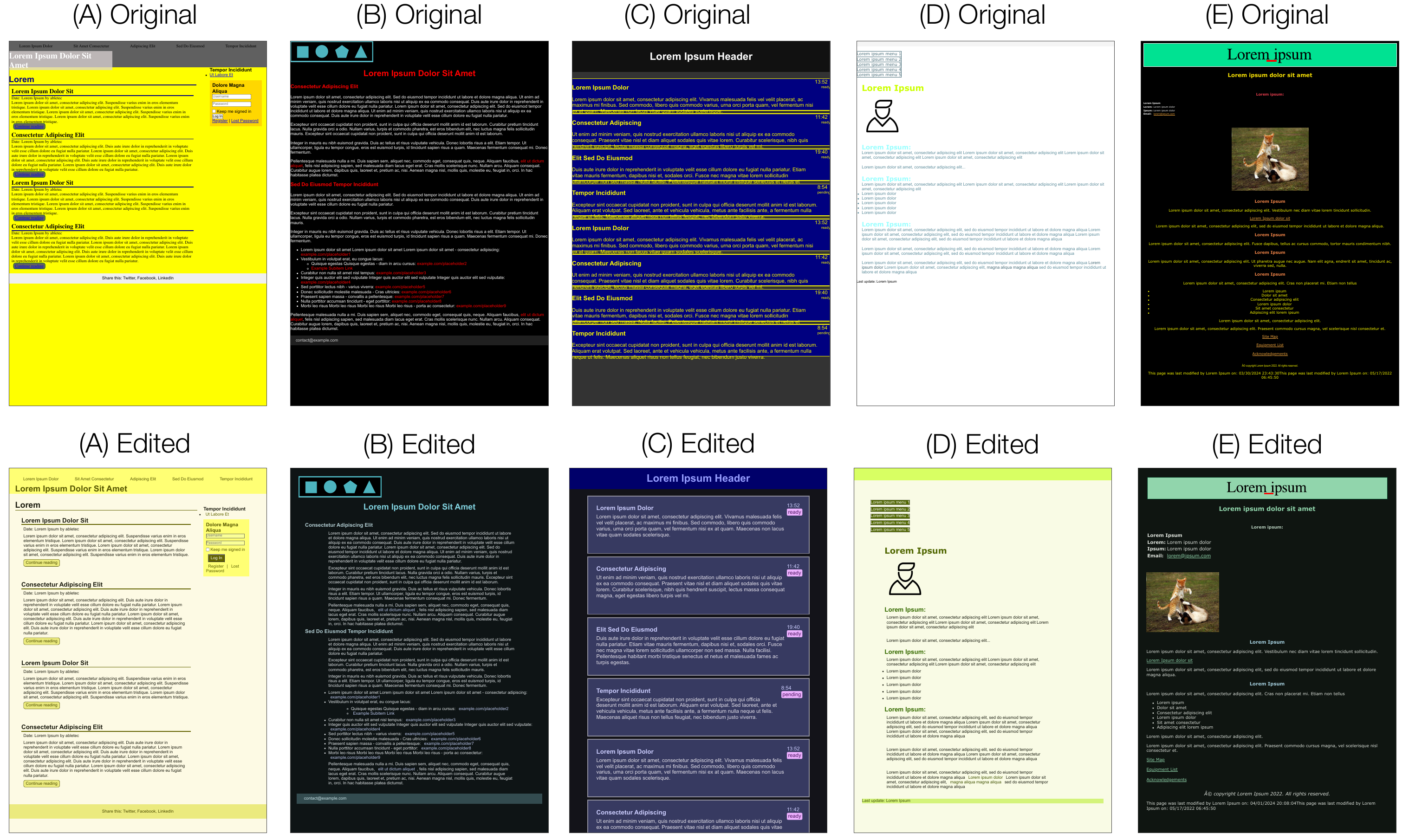}
  \caption{BLV developers' websites edited using \sysname{} (Top: original, Bottom: edited). \sysname{} increased the color contrast of text (A-B, D), introduced more whitespace to separate and align UI elements (A, C-D), and improved color harmony (A-E).}
  \label{fig:website_results}
\end{figure*}

Figure \ref{fig:website_results} shows examples of websites edited using \sysname{}. To generate the results, we selected 5 websites created by BLV developers (collected in \ref{sec:website-analysis}) that cover a diverse range of design issues. 
\revised{We anonymized the selected websites by replacing original text and images with filler text and images, and removing all links.}
We ran \sysname{} on the website recreations to detect design issues and followed the suggestions to resolve the issues. We generated a second report after fixing the initial issues and then resolved the remaining issues.
Overall, \sysname{} increased the color contrast of text (Figure 6A-B, 6D), introduced more whitespace to separate and align UI elements (6A, 6C-6D), and improved color harmony (6A-E).

In Figure~\ref{fig:website_results}A, \sysname{} correctly identified the title linebreak and suggested increasing the surrounding element's width, successfully resolving the linebreak.
\sysname{}'s suggested color scheme also improves overall color harmony and the readability of the header menu and text buttons.
However, on the top-right login box, additional padding resulted in a new linebreak for the text ``Lost Password''. In the second report, \sysname{} suggested increasing the width of the text element to fix the issue, yet this conflicted with the padding value, and the issue remained.
In Figure~\ref{fig:website_results}B, the edited version improves margins and padding around the text.
\sysname{} extracted the main color from the logo image and applied the color scheme that matches the logo.
In Figure~\ref{fig:website_results}C, our system resolves the border/text overlaps.
However, the timecodes on the right still do not center-align as they were optimized to align with the nearest text element (buttons), and the algorithm does not capture alignment between far elements that have other text elements in between.
Figure~\ref{fig:website_results}D improved color contrast between the heading elements and the background. However, the menu items on the top left still lack a margin. 
This results from the original code using a background on hyperlink elements (\smallverb{<a>}) rather than using button elements, and adding a margin did not affect results. 
Also, horizontally aligning the menu items would better utilize the space but providing suggestions for better balance and fixes involving re-ordering HTML elements are currently beyond the scope of \sysname{}.
In Figure~\ref{fig:website_results}E, \sysname{} first left-aligned the bullet text in the original, and then in the second pass aligned other left-aligned paragraph texts and images to improve alignment. The resulting design looks more aligned but less balanced than the developer's original design. \sysname{} revised the color scheme using the original header color, improving the overall harmony. 
\section{User Evaluation}\label{sec:eval_study}
We conducted a within-subjects study to compare \sysname{} with BLV web developers' current web design approaches.
\subsection{Methods}
\subsubsection{Participants}
We recruited 8 BLV web developers who are familiar with HTML and CSS using mailing lists (Table~\ref{tab:participants}). Participants described their vision as totally blind (N=6) or legally blind with light and color perception (N=2) and they all used screen readers. 
4 participants considered themselves to be expert web developers (P4, P10, P11, P12), and created websites for customers or associations (P4, P10, P12) or taught web development (P11). The other 4 participants created personal websites (P1, P13, P14, P15).

\subsubsection{Materials}
We constructed 3 single-page blog websites (Figure~\ref{fig:materials}, W0-W2) using HTML and internal CSS (embedded in HTML as \smallverb{<head>}). The design referenced five blog-type BLV websites collected in the formative study (Section~\ref{sec:website-analysis}). 
We used W0 for the ~\sysname{} tutorial and for collecting ratings of the design comparison descriptions, and W1 and W2 for the editing tasks in the comparison study. W1 and W2 were similar in number of HTML elements (W1: 25, W2: 20), CSS selectors (W1: 12 with 7 tags and 5 classes, W2: 10 with 5 tags and 5 classes), and text length (W1: 377 words, W2: 233 words).

\subsubsection{Procedure}
We conducted a 1.5-hour remote study using Zoom. We first asked background questions about participants' prior web design experiences. We then asked participants to read three versions of the design comparison generated with W0: 
(1) guideline-only, (2) guideline and reference website comparison, and (3) guideline, reference website, and trends of 20 blog websites.
Participants rated the comparison versions on a 7-point Likert scale to assess ease of use and usefulness for web design.
Next, we provided a brief walkthrough of installing the extension and gave a 15-minute tutorial on the full ~\sysname{} system with W0. Then, participants edited two websites (W1, W2), one using their current approaches (\textit{e.g.} screen reader add-ons, search engine) and one using ~\sysname{}. We allowed participants to use the code editor of their choice for both conditions. The order of the websites and the tools used were counterbalanced and randomly assigned to participants. For each condition, participants were given 20 minutes to make edits. After each editing task, we conducted a post-stimulus survey using selected cognitive load measures from NASA-TLX (\textit{Mental Demand}, ~\textit{Frustration}, ~\textit{Effort}, ~\textit{Performance})~\cite{hart1988development}, as well as task-focused measures (\textit{Usefulness of the tool(s) in identifying design issues}, ~\textit{Usefulness of the tool(s) in resolving design issues}, and ~\textit{Confidence in the final output}). All ratings were on a 7-point Likert scale. Finally, we conducted a semi-structured interview to understand participants' strategies for using the tools, as well as the limitations and opportunities of ~\sysname{}.

We analyzed the video recordings, transcript, survey responses, and the participants' final website outputs. 
Using the Firebase database, we logged participants' interactions with the system (\textit{e.g.,} \textit{Design Comparison} options setting that users selected, and the analysis results in each attempt).

\subsection{Findings}
\subsubsection{Using Design Comparisons}
\begin{figure}
  \centering
  \includegraphics[width=3.33in]{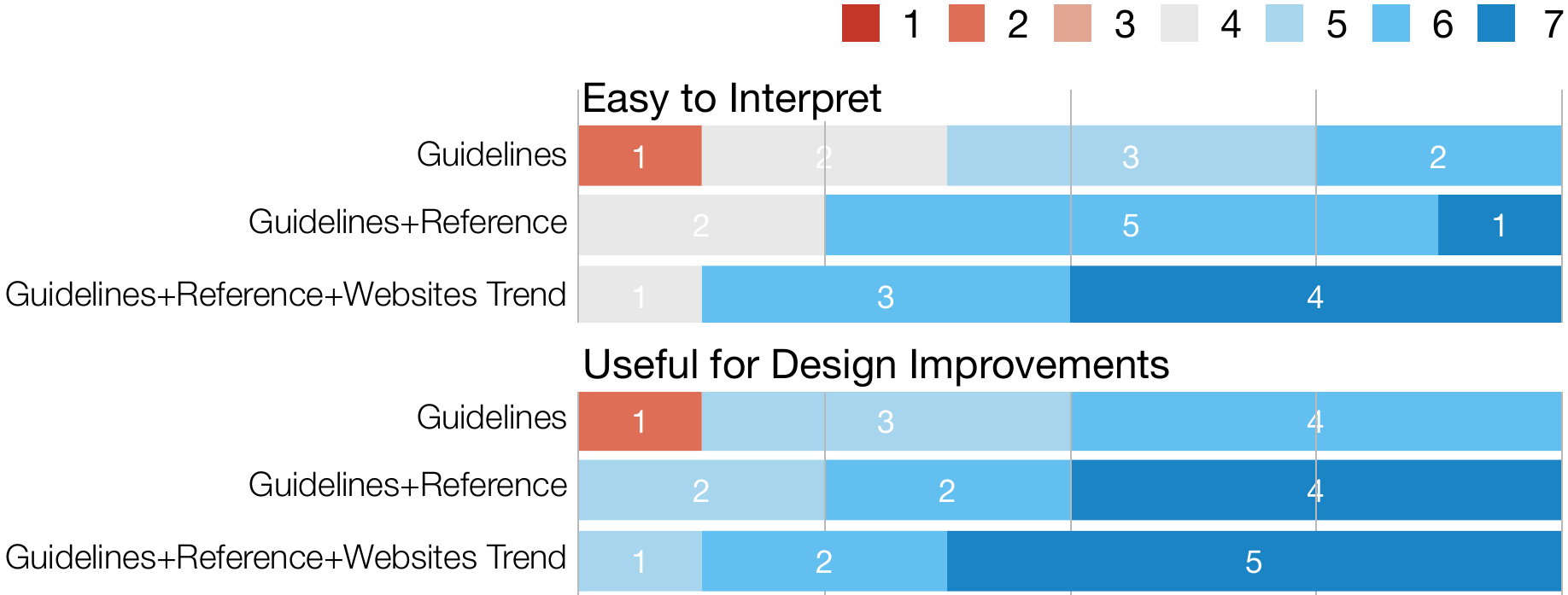}
  \caption{We report the distribution of ratings of different design comparisons: (1) visual design guidelines (2) a reference website, and (3) and a trend of multiple websites.}
  \label{fig:comparison_rating}
\end{figure}

Figure~\ref{fig:comparison_rating} shows the ratings of three different versions of design comparisons. We reveal how participants used the three types of design comparisons in the website editing tasks.

\ipstart{Guidelines support interpreting suggestions}
Participants mentioned the benefit of guidelines for learning design concepts (P1, P5, P12). 
P12 said ~\textit{``Guidelines help beginning web designers.''} Similarly, P5 stated ~\textit{``I actually already learned a few design rules from using this [\sysname{}] for a few minutes!''}
They also highlighted the importance of guidelines when interpreting the styles of a reference website or website trends. P10 stated ~\textit{``First, I should look at guidelines and then the other references. I can’t interpret these without guidelines. Foundational information is defined in the guidelines.''}


\ipstart{Reference website provides contextualized suggestions}
When choosing a reference website for the editing task, participants selected websites that they often use (P14-P15), have many visitors (P1, P4, P10, P12-P15), belong to someone who they know (P10-P11), and are screen reader accessible (P15).

Participants highlighted that comparison with their selected reference website can add clarity to the guidelines. For instance, P10 said ~\textit{``The reference website gives me a better idea of how guidelines should be actually implemented.''} P1 noted that the reference website provides contextualized feedback. He noted ~\textit{``Guidelines provide generalized suggestions that can be for any website, but this one [reference website] provides personalized suggestions for my context, professional blog in this case.''} 
Yet, during the website editing task, two of the participants mentioned the difficulty of selecting a reference website to use without access to the visual design (P1, P4). In addition, P13 raised copyright concerns when following the reference website. She noted ~\textit{``I won't copy all the styles because I still want mine to look different. For lower-level design choices like font sizes or margins, I can copy them. For high-level design choices, I want to do those on my own layout or colors. I still want to keep my creativity.''} 
During the website editing task, 3 participants noticed that suggestions from guidelines were different than what is used in the reference website. For instance, the guidelines suggested 16-20\smallverb{px} for the body text, the reference website that P4 selected used fonts bigger than 20\smallverb{px}.

\ipstart{Website trends increase confidence}
Participants said understanding the trends in multiple websites increases their confidence (P1, P4, P10-P12, P15). P15 stated ~\textit{``If most other websites are using certain styles, I'll be more confident in following that trend.''}
P1 also noted that website trends can save time in reviewing other designs and wanted to use it in earlier stages of the design process ~\textit{``Before I start designing my website, I can check these trends. Just like people looking at fashion magazines before shopping for clothes.''} P1 described his priority in different suggestions ~\textit{``If there are conflicts between the guidelines and the trends of multiple websites, I’ll follow the trend. Because that is the reality.''}

While ~\sysname{} provided 5 different website categories (Blog, Tutorial, Personal Website, Organization Website, and News/Magazine), participants expressed that they wanted more control over what constitutes each category (P1, P3, P12). P1 mentioned that he wants to curate a dataset to find the right trend for specific contexts ~\textit{``Design preferences may vary depending on different cultures and scenarios. Popular font families in the U.S. might not be in China.''} 
P12 noted ~\textit{``It is important to regularly update the websites to reflect the latest trend.''}


\begin{figure}[t]
  \centering
  \includegraphics[width=3in]{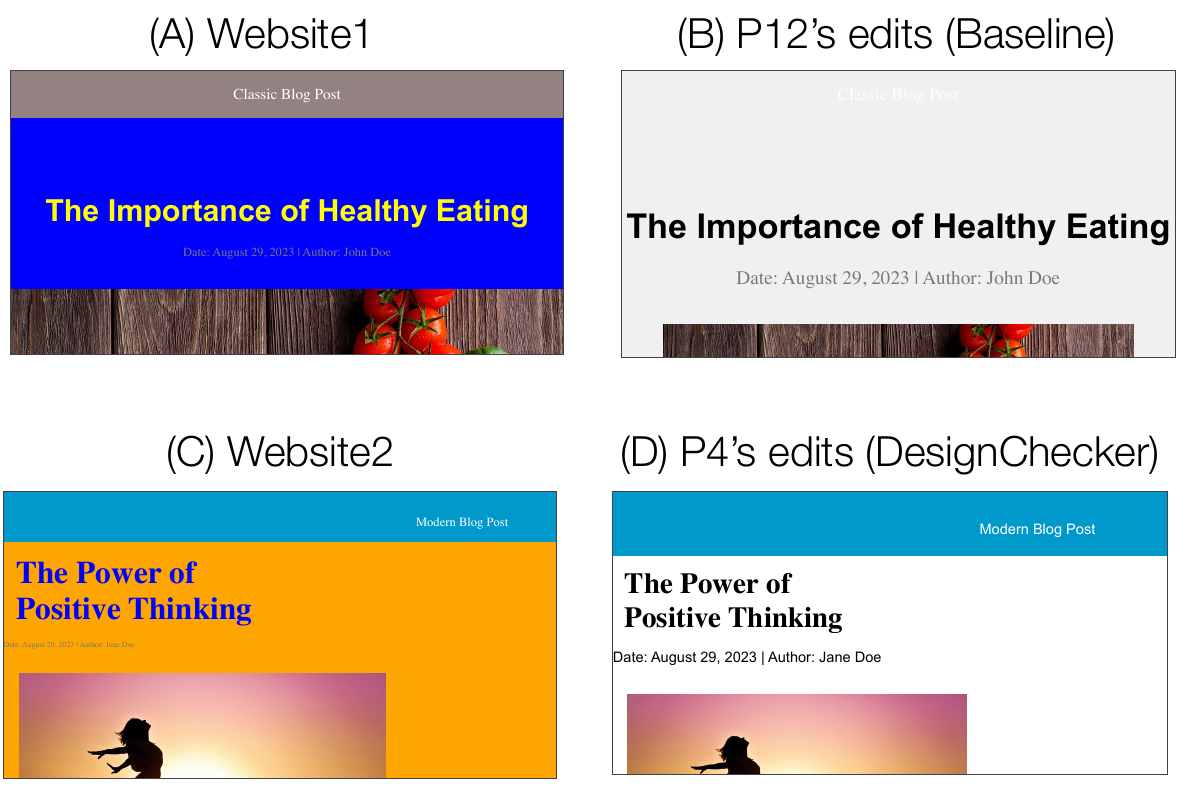}
  \caption{Example websites from the study: Original design of W1 (A), W1 edited by P12 with baseline (B), Original design of W2 (C), W2 edited by P4 with DesignChecker (D).}
  \label{fig:example_websites_w1}
\end{figure}


\subsubsection{\sysname{} vs. existing workflows}
\begin{figure*}[t]
  \centering
  \includegraphics[width=0.95\textwidth]{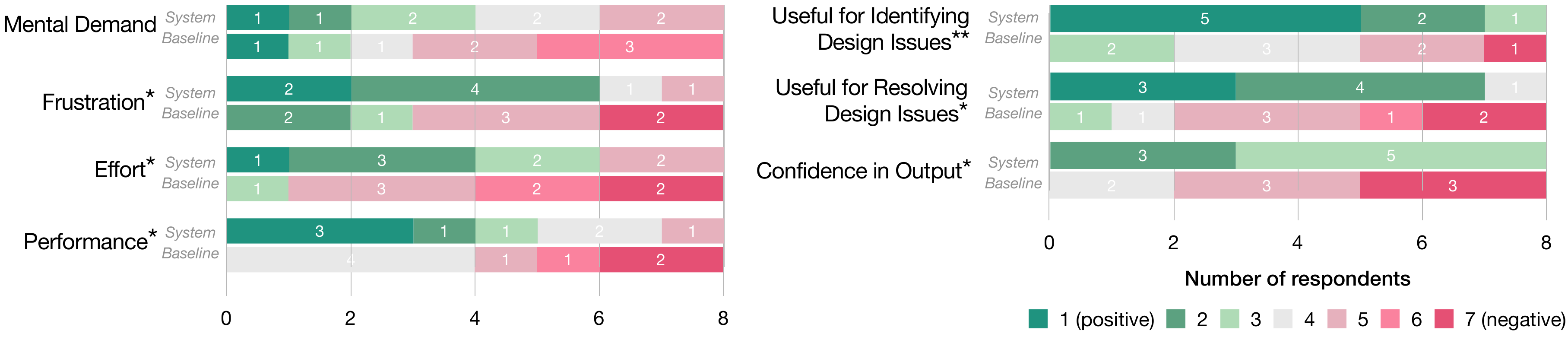}
  \caption{Distribution of the rating scores for the participants' personal editing tools and ~\sysname{} (1 = low, 7 = high). Note that a lower value indicates positive feedback and vice versa. 
  {The asterisks indicate the statistical significance as a result of Wilcoxon text} (\textit{p} < 0.05 is marked with * and ~\textit{p} < 0.01 is marked with **).}
  \label{fig:eval_likert_scale}
\end{figure*}

In the comparison study, participants edited two websites, one using ~\sysname{} and one with current workflows (\autoref{fig:eval_likert_scale}). All participants expressed that they wanted to use ~\sysname{} in the future for web designing. Overall, participants rated ~\sysname{} as requiring significantly less frustration ($\mu$=4.50, $\sigma$=2.00 vs. $\mu$=2.38, $\sigma$=1.41; $Z$=-1.70, $p$<0.05) and effort ($\mu$=5.50, $\sigma$=1.31 vs. $\mu$=2.88, $\sigma$=1.46; $Z$=X, $p$<0.05), while achieving significantly higher performance ($\mu$=2.88, $\sigma$=1.36 vs. $\mu$=5.38, $\sigma$=1.60; $Z$=2.18, $p$<0.05). Differences in mental demand were not significant~\footnote{We used the Wilcoxon Signed Rank test to perform significance testing.}. 

\ipstart{Reviewing and Identifying Design Issues}
\revised{Before running DesignChecker, 7 participants used a screen reader to access the rendered webpage to understand its content and layout. 4 participants reviewed the HTML/CSS code in their editor (\textit{e.g.,} VS Code). After running DesignChecker, 6 participants reviewed feedback in the presented order, 2 participants (P13, P15) reviewed in order of perceived severity (\textit{e.g.,} impacting many elements), and P1 followed his personal priorities. He noted ~\textit{``I'm checking out the colors here first because I know that is the hardest issue that I can't solve by myself.''} While DesignChecker provides brief explanations behind its feedback, 3 participants also clicked the detailed guidelines to gain foundational design knowledge (P1, P10-11). For instance, P1 learned why saturated colors are unsuitable for backgrounds while reading about the definition of saturation.
Participants mentioned that \sysname{}'s list of \textit{passes} were also useful for learning (P13, P14). P14 noted ~\textit{``Providing correct elements and why they are okay is good for learning purposes. It reinforces what I'm doing right.''}}

In the baseline condition, participants adopted multiple strategies to review the web design -- a search engine to check the color names (P1, P12) and a large-language model (ChatGPT) to describe the color codes, evaluate a design choice in CSS, or recommend suggestions (P10, P14). 
Yet, the reviews from these tools lack a comprehensive view of the website and offer generic suggestions which often lead to additional issues. 
Also, reviews occurred only when a design choice was questioned by participants, with the majority of issues unnoticed. 
As a result, 3 participants using their current approaches finished editing early  (P4, P11, P15, ~\autoref{tab:edit_results}), while all participants using ~\sysname{} spent 20 minutes allotted for editing the websites. P15, who only made 2 changes during the task said ~\textit{``There’s nothing much that I can edit further. I know something's wrong but I don't know what they are.''}

Participants found ~\sysname{} to be significantly more useful in identifying design issues than their existing workflows ($\mu$=3.63, $\sigma$=1.30 vs. $\mu$=6.50, $\sigma$=0.76; $Z$=2.49, $p$<0.01) (Figure~\ref{fig:eval_likert_scale}). P12 noted ~\textit{``This is just like an accessibility checker for sighted people, but for blind people!''} P15 particularly liked the comprehensive reviews that ~\sysname{} provides ~\textit{``Average people won’t be able to review in this much detail. They always just say ``this doesn’t look good enough.''} ~\sysname{}'s design suggestions provided opportunities for gaining new CSS knowledge as P13 described ~\textit{``I gained new knowledge about the defaults in CSS - I didn't think that not specifying a font-family will use [Times New Roman] and that would be a problem.''}

Participants suggested improvements for ~\sysname{}. P12 wished that the severity of each design issue was communicated. She explained ~\textit{``I want the presenting order [of issues] to be from most critical to least critical. White text on a yellow background is a must-fix, but changing to a modern font family is optional.''} Also, P14 wanted the option to group the issues by elements instead of the design category. He described ~\textit{``When I'm actually editing, I want to minimize the back and forth navigation in the code editor. So issues listed per element will be easier. Then, I can also decide to skip reviewing less important elements -- because I care more about the main article and less about the footer.''}

\ipstart{Resolving Design Issues}
~\autoref{tab:edit_results} summarizes the number of edits and the number of issues resolved by each participant in both conditions. Using ~\sysname{} participants made a significantly higher number of edits ($\mu$=4.25, $\sigma$=2.12 vs. $\mu$=7.75, $\sigma$=2.71; $Z$=2.03, $p$<0.05) (Figure~\ref{fig:eval_likert_scale}), and resolved more issues ($\mu$=2.75, $\sigma$=1.83 vs. $\mu$=6.63, $\sigma$=2.77; $Z$=2.04, $p$<0.05) (Figure~\ref{fig:eval_likert_scale}).

\revised{Participants resolving issues with DesignChecker used arrow keys and search (Ctrl-F) in their editor to locate the CSS class that DesignChecker listed with an issue. 5 participants copied DesignChecker's recommended fixes into their editor, and 3 participants manually typed the CSS changes. P10 avoided manual typing as he could not notice typos. Most participants followed DesignChecker's guideline suggestions, but P11 selected an intermediate font size between the guidelines and his reference website. Upon changing the code, P14 further navigated the page with a screen reader to check for compile issues.}

In the baseline condition, 2 participants resolved only text-related issues (P11, P15). P15 noted ~\textit{``I only changed the font sizes because they're the only things I know for sure!''} 
P1 noted the difficulty of planning the edits without ~\sysname{}. He stated ~\textit{``I don't have the big picture of what's wrong with my website and just fixing individual style classes one by one.''}
When participants using the baseline attempted to resolve an issue, it often led to additional design issues. For instance, when P12 changed the background color of the header to white to match the article's background, the white text on the header became invisible (Figure~\ref{fig:example_websites_w1}B).


Participants perceived ~\sysname{} significantly more useful for resolving design issues than their existing tools ($\mu$=2.75, $\sigma$=1.39 vs. $\mu$=6.13, $\sigma$=1.00; $Z$=2.32, $p$<0.05) (Figure~\ref{fig:eval_likert_scale}). 
When following design suggestions, 3 participants also referred to the design comparison to decide what CSS changes to make (\textit{e.g.,} P11 referred to the most popular font family in the blog dataset websites). 
P10 highlighted the benefit of the contextual suggestion that ~\sysname{} provided. He shared ~\textit{``All suggestions contextual to my codes are consolidated in one place, so extremely more useful than chatGPT's general descriptions.''} 
P12, a professional web developer enjoyed being able to choose which recommendations to implement. She said ~\textit{``Automating all the suggestions would be a bad idea because it might change into something that I don't really like. If it has color issues, I have the flexibility to modify either the foreground or background.''} In contrast, P1, who developed websites as a hobby wanted the option to auto-resolve all the issues.
Because \sysname{} provided feedback per group of HTML elements that share the same style classes, participants could easily search (ctrl-F) for the CSS property names in the code editor and make fixes (P15, P12).

With ~\sysname{}, participants were significantly more confident with their final web design ($\mu$=2.5, $\sigma$=1.31 vs. $\mu$=5.38, $\sigma$=0.52; $Z$=2.47, $p$<0.01) (Figure~\ref{fig:eval_likert_scale}). 
After making edits, ~\sysname{}'s summary of changes helped users to confirm that their edits to the code actually resolved the design issues and improved the design as P15 noted ~\textit{``If I run it again [after making design edits], and it [\sysname{}] says my design improved, I'll be so confident!''}

\begin{table}[h]
\small\sffamily\def\arraystretch{1.2}\setlength{\tabcolsep}{0.4em}
    \centering
    \begin{tabular}{c|cc|cc}
        \toprule
       \multirow{2}{*}{PID}  & \multicolumn{2}{c|}{\# of Design Edits} & \multicolumn{2}{c}{\# of Resolved Issues}  \\
       \cline{2-5}
        & Baseline & \sysname{} & Baseline & \sysname{}  \\
       \midrule
        P1  & 7 (1t, 2l, 2c, 1o) & 12 (3t, 7l, 2c) & 4 (1t, 2l, 1c) & 9 (3t, 5l, 1c)\\
        P4 & 1 (1c) & 8 (5t, 3c) & 0 & 7 (5t, 2c) \\
        P10  & 7 (2t, 1l, 4c) & 5 (3t, 2c) & 6 (2t, 1l, 3c) & 5 (3t, 2c) \\
        P11  & 4 (4t) & 7 (2t, 1l, 3c, 1o) & 2 (2t) & 5 (2t, 1l, 2c)\\
        P12  & 5 (2t, 3c) & 4 (2t, 2c) & 2 (2t) & 4 (2t, 2c) \\
        P13 & 4 (2t, 2c) & 8 (2t, 1l, 5c) & 2 (2t) & 6 (2t, 1l, 3c) \\
        P14 & 4 (1t, 2l, 1c) & 11 (6t, 2l, 3c) & 4 (1t, 2l, 1c) & 9 (5t, 2l, 2c)  \\
        P15 & 2 (2t) & 7 (5t, 1l, 1o) & 2 (2t) & 6 (5t, 1l) \\
        \midrule
         \multicolumn{1}{c|}{\textbf{$\mu$}} & 4.25 & 7.75 & 2.75 & 6.63 \\
        \midrule
         \multicolumn{1}{c|}{\textbf{$\sigma$}} & 2.12 & 2.71 & 1.83 & 2.77 \\
        \bottomrule
    \end{tabular}
    \caption{The number of design edits and the issues resolved by participants using their current tools (Baseline) and ~\sysname{} (t: text, l: layout, c: color, o: others). Both the number of design edits and the issues resolved showed the statistical significance with Wilcoxon test (\textit{p} < 0.05)}
     
    \label{tab:edit_results}
\end{table}

\section{Discussion}
We reflect on our findings from the 
design and evaluation of \sysname{} and discuss the lessons and future opportunities. 

\ipstart{Evaluation by Design Comparison}
\sysname{} lets users compare their web design with multiple resources (\textbf{D3}) that have different advantages as noted by participants. The guidelines helped them interpret the suggestions learn design concepts. Comparing with a reference website or the trend of multiple websites enabled context-specific suggestions. 
By comparing with multiple resources, participants could learn from conflicting suggestions (\textit{e.g.,} a reference website using a different font size than recommended by guidelines) and find alternative design choices rather than copying a single website's design. 
Yet, it remains challenging for BLV users to search for references (\textbf{D2}). Future work can explore generation of a high-level design descriptions to facilitate accessible design search. 
We can also extend ~\sysname{} to let users manually curate websites for learning trends based on various criteria, such as date posted or number of visitors as noted by participants. 


\ipstart{Editing Based on Design Suggestions}
To support BLV developers in identifying and resolving design issues (\textbf{D4}, \textbf{D5}), \sysname{} detects common design issues in BLV websites and suggests fixes. 
By providing feedback per group of HTML elements that share the same style classes, ~\sysname{} enabled users to remember the usage of the elements, quickly navigate to the relevant class, and decide what to change.
Study participants also mentioned that when they reviewed the summary of changes and noticed that issues were resolved (\textbf{D6}), they were more confident with the results. 

~\sysname{} did not provide an automated fix to encourage users to review the suggestions and make decisions as informed by our formative study. Yet, there are benefits of auto-fix in certain contexts: when the users need to quickly implement the website, when they are less interested in learning design concepts, or when they have used the system for a long time and gained trust so that they do not need to review the suggestions. To help users prioritize which issues to address, future versions of ~\sysname{} can sort design issues based on their prominence (\textit{e.g.,} issues affecting prominent elements like headers) or severity (\textit{e.g.,} undiscernible text due to extremely low contrast).

\ipstart{Learning about Design}
We need more accessible resources for learning visual design (\textbf{D1})~\cite{kearney2021accessible, pandey2021understanding}. While ~\sysname{} was not primarily designed to teach design concepts, our study participants found it helpful for learning visual design. It explains guidelines and demonstrates their application on other websites, thereby improving users' understanding of design concepts and trends.
Future work can improve access to various resources for learning design by providing detailed descriptions for images used in tutorials (\textit{e.g.,} describing changes after a design effect is applied~\cite{huh2023genassist}) or making tutorial videos accessible~\cite{liu2022crossa11y}. 
For intuitive learning of spatial concepts, we can also explore non-visual tutorials using audio~\cite{saha2023tutoria11y} or touch~\cite{potluri2021examining, kearney2021accessible}.
While sighted people can naturally learn about designs as they visit others' websites and notice common patterns, not all of the tacit knowledge in design can be identified in code or articulated as a description. Future research is needed to explore accessible in-situ learning and delivery of tacit knowledge in design.

\ipstart{Integration into Current and Future Workflows}
We built ~\sysname{} as a browser extension that presents feedback and suggestions as a screen-reader-accessible HTML. 
We reflect on how ~\sysname{} can be better integrated into users' current tools. By introducing a screen reader add-on~\cite{NVDAtoolkit}, ~\sysname{} can provide notifications for design issues as users navigate through the website with a screen reader. Additionally, an extension for code editors can assist users in promptly identifying and addressing issues while implementing the code. 
As our formative study revealed, BLV developers with remaining vision often adapted the design to make the reviewing easier (\textit{e.g.,} using dark borders or high contrast background). To support low-vision users, we can explore visual feedback to better communicate design issues (\textit{e.g.,} overlaying visual effects to highlight elements with design issues). 
There is also an opportunity to utilize built-in components from UI frameworks (\textit{e.g.,} Bootstrap) which are preferred by developers less interested in customization~\cite{pandey2022accessibility}. \sysname{} can suggest a built-in component with similar functionalities (\textit{e.g.,} search field, menu button) or support better integration of the components into the current design (\textit{e.g.,} adjusting the color of UI components to match the current color scheme). 
\revised{In the future, we expect ~\sysname{} to complement other accessibility tools, rather than function as a standalone tool. For example, with a tactile display, users can feel where and how the text elements are displayed but ~\sysname{} can further describe that center-aligned paragraphs impact readability. A color sonification tool can communicate that 15 different colors are used, but ~\sysname{} can explain that using too many colors can lead to a visual distraction. Our future work will explore how to best convey design feedback across multiple modalities.}

\ipstart{Support for Other Design Aspects}
~\sysname{} currently supports users in identifying design issues related to text, layout, and colors.
\revised{Yet, some challenges remain uncovered with our tool, such as verifying that images are rendered with the right size and ratio~\cite{kearneyaccessible, zhang2023a11yboard}, or evaluating the density or symmetry of the design~\cite{lee2020guicomp}.
A future iteration of ~\sysname{} can provide feedback on broader design aspects including balance or density and predict where website visitors will focus on using attention heatmap and visual saliency~\cite{bylinskii2017learning}.}
Also, while ~\sysname{}  design suggestions that involved CSS changes, future work will consider re-designing the layout that involves HTML changes and creating interactive or responsive designs with JS.


\ipstart{Generalizability of \sysname{} for Accessible Authoring}
We addressed accessible web development using HTML and CSS and did not explore other types of UI development (\textit{e.g.,} mobile application, game, or VR interface), leaving opportunities for future work. 
Such support should consider the unique nature of each application -- for instance, mobile UI development can leverage touch exploration that can verify the presence of UI elements~\cite{potluri2019multi}.
We can also extend \sysname{}'s approaches for visual design feedback and support accessible authoring beyond programming contexts -- slide design~\cite{peng2022diffscriber}, art~\cite{huh2023genassist}, photo, video~\cite{huh2023avscript}, or fashion~\cite{burton2011fashion}. For instance, we will explore how ~\textit{Design Suggestions} can be used to inform slide authors about readability issues, and how ~\textit{Design Comparison} can describe fashion trends and give suggestions on selecting matching clothes.

\ipstart{Beyond Checker}
As a ~\textit{checker}, our system evaluates a web design based on ~\textit{design norms} and guides users to follow them. While this approach is useful for identifying common design issues with rationales, future work may explore how accessible web design can be more interactive. For instance, conversational assistants can support users in asking additional questions not covered in the pre-set metrics (\textit{e.g., ``Does my navigation bar stand out?''}) or requesting alternative suggestions (\textit{e.g., ``I want brighter color suggestions.''}). 
Additionally, systems powered by code generation AI (\textit{e.g.,} copilot~\cite{Copilot}) can assist in earlier stages of web design, helping users to create the foundational layout or achieve specific design goals (\textit{e.g., ``I want to highlight the top banner with warm colors.''}) 
\revised{However, it is important to consider the drawbacks of code-generation approaches, as they can lead to overly common designs or resemble designs that are copyrighted.
While our formative study findings indicate that BLV developers work on styles in later stages, readily available feedback with ~\sysname{} can shift their workflow to consider design from the beginning.
Future work can explore how adopting ~\sysname{} in earlier stages impacts user processes (\textit{e.g.,} less repeated design issues).}

\ipstart{\sysname{} and Visual Communication}
 We live in a world that prioritizes visual communication. As noted by our participants, bad design might influence their work to be perceived as less professional, and we developed ~\sysname{} to support BLV people to gain more confidence in communication through websites. However, by setting sighted individuals' expectations and standards of visual design with BLV developers, ~\sysname{} still runs the risks of putting a burden on BLV developers to follow them. We need further technical and social initiatives to improve the accessibility of current websites and make design tools more accessible. 





\ipstart{Limitations}
Our system design and study setup have several limitations. First, our system does not cover all design guidelines, but we prioritize design issues commonly identified in BLV developers' websites.
To provide the design trends, we cover 5 categories that BLV developers often create or visit (\textit{e.g.,} blogs, personal websites). 
While current \sysname{} may not capture all website categories and styles, future iterations of \sysname{} can support the update of guidelines and the set of websites to reflect the latest design trend and let users manually curate their own dataset.


Our pipeline is powered by algorithms and AI models and is prone to errors. 
For instance, OCR may be incorrect when the text and background have a similar color, and GPT-4 may output inconsistent color namings.
We did not explore how users perceive and react to errors as no study participant mentioned noticing incorrect feedback. 
GPT-4 also has a limit for input tokens (8.1k) and the pipeline cannot process websites with extensive code spanning multiple CSS files.
In the future, large multimodal models that take visual input (\textit{e.g.,} GPT-4V) can be integrated to improve the efficiency and performance of the pipeline.

In our controlled user study, participants edited the example websites with a time constraint (40 minutes for 2 websites) to balance realism and cognitive demand. As a result, we did not explore how participants use ~\sysname{} to edit their own websites or edit websites with more time provided (\textit{e.g.}, many participants wanted to address all of the identified issues).



\section{Conclusion}
We present ~\sysname{}, a system that supports accessible web design by identifying visual design issues and providing improvement suggestions. With ~\textit{Design Comparison}, users can compare their current web design to visual design guidelines, a reference website of their choice, or a trend of multiple websites. With ~\textit{Design Suggestions}, users can identify the specific HTML elements with design issues and follow suggestions to make CSS changes. Our comparison study with 8 BLV developers demonstrates the effectiveness of ~\sysname{}. 
Our work identifies a latent demand for accessible design tools and we hope this research will catalyze more work in supporting people with disabilities to express their creativity through design.
\begin{acks}
We thank our study participants for their valuable feedback that informed the design and revisions of \sysname{}.
\end{acks}

\bibliographystyle{ACM-Reference-Format}
\bibliography{sample-base}
\clearpage
\onecolumn
\appendix

\section{Website Design Analysis}~\label{apndx:website_analysis}
First, the researchers reviewed design principles from Luther et al.~\cite{luther2014crowdcrit, luther2015structuring} that compiles a holistic set of 70 visual design critique statements from design textbooks (\textit{e.g., A primer of visual literacy}~\cite{dondis1974primer}). 
One researcher open-coded the websites by referencing the design critique statements and searched for relevant web-specific design guidelines for each issue type. The design guidelines were chosen based on Sadler's criteria~\cite{sadler1989formative} for good feedback: specific, conceptual, and actionable. 
Then, the other researcher reviewed the same data for conflict resolution and validation. Two researchers discussed it further to group the guidelines into high-level categories summarized in \autoref{tab:form-categories}.
Finally, using the design categories, they counted the number of issues for each website in the dataset. We report the number of identified design issues in \autoref{tab:form-analysis}. Note that when assessing color harmony, we did not count the number of elements with issues but instead rated holistically (as the issue is determined by a website as a whole rather than from individual elements). 

\begin{table*}[!htbp]
\small\sffamily\def\arraystretch{0.6}\setlength{\tabcolsep}{0.4em}
\centering
\begin{tabular}{l l p{8cm} p{2cm}} 
\toprule
Category & Sub-category & Guidelines & CSS Properties \\ 
\midrule
\multirow{5}{*}{Text} 
 & Legibility
 & Body fonts should be 16px or above. Title fonts should be 20px or above.~\cite{digitalgov2023, learnuidesign} & \smallverb{font-size} \\ 
 & \multirow{2}{*}{Readability} 
 & Sans-serif fonts are generally considered more legible than Serif fonts. Decorative and narrow fonts should be used for headlines and decorative texts only.~\cite{psuaccessibility, uncgaccessibility, uxplanet2023} & \smallverb{font-family} \\ 
 & & Optimal line length for body text is 50–75 characters. Avoid line breaks in title texts.~\cite{baymard2023, creativepro2023, fonts2023} & \smallverb{font-size, width} \\ 
 & & Line spacing should be at least 1.5 times the font size within paragraphs~\cite{baymard2023, uxdesigncc2023} & \smallverb{line-spacing} \\
\midrule
\multirow{3}{*}{Layout} 
 & Spacing & UI Elements should have a minimum of 8px of margin between them. Containing elements need at least 24px of padding.  ~\cite{uxengineer2023, materialio2023} & \smallverb{margin, padding} \\ 
 & Spatial Alignment & Close elements should be aligned according to their edges or central axis (horizontal or vertical).~\cite{uxknowledgebase2023, aelaschool2023} & \smallverb{width}, \smallverb{height}, \smallverb{margin}, \smallverb{padding}, \smallverb{justify-content} \\ 
& Textual Alignment & Long text should be left-aligned (use center for shorter texts \textit{e.g.,} headlines.) Never center-align bullet texts.~\cite{felterunfiltered2023} & \smallverb{text-align, justify-content, display} \\ 
\midrule
\multirow{2}{*}{Color} 
 & Color Contrast & Provide ample lightness contrast between text and background. Avoid using high-saturated background colors. ~\cite{readtech2023, w3c2023, uxmovement2023} & \smallverb{background-color, color} \\
 & Color Harmony & Consider color harmony and choose a set of colors that work well together.~\cite{nngroup2023, uxmovement2023} & \smallverb{background-color, border-color, color}\\ 
\bottomrule
\end{tabular}%
 \caption{We collected and grouped web design guidelines related to common design issues found in the website analysis, which informed the design of ~\sysname{}. In the ~\sysname{} interface, we modified the sub-categories names to be more intuitive for BLV people and easy to map with the corresponding CSS properties (~\textit{e.g.,} ~\textit{font family} or ~\textit{line-spacing} instead of ~\textit{readability}).}
  \label{tab:form-categories}
\end{table*}
\clearpage

\section{STUDY PARTICIPANTS DEMOGRAPHICS}

\begin{table*}[htbp!]
\small\sffamily\def\arraystretch{1}\setlength{\tabcolsep}{0.5em}
    \centering
    \begin{tabular}{lllllll}
        \toprule
       PID  & Gender & Age & Vision & Onset & Job & Websites Created \\
       \midrule
        P1{$^{*}$} & Male & 37 & Totally blind & Congenital & Graduate student & Blog \\ 
        P2 & Male & 63 & Totally blind & Acquired & Corporate trainer & Association website \\ 
        P3 & Female & 23 & Legally blind & Congenital & Web accessibility consultant & Blog, Association website, Personal website, \\ 
        P4{$^{*}$} & Male & 28 & Legally blind & Congenital & Graduate student & Tutorial, Personal website, Association website \\ 
        P5 & Male & 50 & Totally blind & Acquired & Software engineer &  Tutorial, Blog, Personal website, Product website\\ 
        P6 & Male & 46 & Legally blind & Congenital & Software engineer & Association website, Personal website\\ 
        P7 & Male & 47 & Legally blind & Acquired & Software engineer & Tutorial, E-commerce, Associaiton website\\ 
        P8 & Male & 58 & Totally blind & Acquired & Software engineer & Tutorial\\ 
        P9 & Male & 44 & Totally blind & Congenital & Information security consultant & Blog\\ 
        P10 & Male & 29 & Totall blind & Congenital & Accessibility consultant & Product website, Chrome extensions\\ 
        P11 & Male  & 51 & Totally blind & Congenital & Lecturer & Personal website, Course website \\ 
        P12 & Female & 71 & Totally blind & Acquired & Web developer & Blog, Tutorial\\ 
        P13 & Female & 20 & Legally blind & Acquired & Undergraduate student & Blog\\ 
        P14 & Male & 30 & Totally blind & Acquired & Graduate student & Personal website\\ 
        P15 & Male & 22 & Totally blind & Congenital & Accessibility consultant & Personal website, Podcast website\\ 

        \bottomrule
    \end{tabular}
    \caption{Demographics of study participants (Formative study: P1-P9, Evaluation study: P1, P4, P10-P15, {$^{*}$}P1 and P4 participated in both studies.). All participants had web development experience and used screen readers (NVDA and JAWS).} 
    \label{tab:participants}
\end{table*}

\section{GPT-4 PROMPTS }
\sysname{}'s pipeline is powered by GPT-4 to identify design issues and provide suggestions for font family and colors (Section~\ref{pipeline_design_audit}). We used the following prompts with temperature of 0.2:

\subsection{Font Family Suggestions}

\begin{table*}[!htbp]
\scriptsize
\resizebox{\textwidth}{!}{%
\begin{tabular}{@{}p{15cm}@{}}
\toprule
Check the font families of the website and give suggestions. Sans Serif is preferred to Serif for better readability. The response should be in a single array with three elements.\\ 
First element: string, one-sentence summary of whether the fonts used have good readability (Sans-Serif) and are visually appealing. \\
Second element: object, details of elements that need to change the font family. Each key is the group name. If the current one is okay, you don't need to include it in the object. Don't be abstract and suggest what can directly be applied to CSS. \\
Third element: object, details of elements to change the font family.\\
Example response: \\\relax
[``The use of Georgia for paragraphs, Times New Roman for headers, and Arial for buttons offers reasonable readability but may lack visual consistency due to the mix of serif and sans-serif fonts.''
{{``p-title'': ``To maintain a more consistent look, you could switch to a sans-serif font like Arial or a modern serif like Georgia to match the body text.''}, ...}, {{``p'': ``The readability of Arial is good.''}, ...}]\\
\bottomrule
\end{tabular}
}
\caption{A full example of prompt used in font family suggestion.}
\end{table*}

\subsection{Current Color Scheme Summary}~\label{apndx:color_scheme_prompt}
\begin{table*}[!htbp]
\scriptsize
\resizebox{\textwidth}{!}{%
\begin{tabular}{@{}p{15cm}@{}}
\toprule
Check how consistent and harmonious the website's colors used are.\\
The response should be in a single array with two elements. First: string, a summary of the current color scheme. Second: object, current color scheme details.\\
Example response: \\\relax
["The color palette used in the website primarily consists of ...", 
{"Background Color": "Soft Pastel Blue (rgb(168, 180, 255), rgb(194, 206, 255)) dominates interactive elements like 'button' and 'span'. ...", "Border Color": ...}]\\
\bottomrule
\end{tabular}
}
\caption{A full example of prompt used to summarize the current color scheme of a website.}
\end{table*}

\clearpage

\subsection{Color Suggestions}
\begin{table*}[!htbp]
\scriptsize
\resizebox{\textwidth}{!}{%
\begin{tabular}{@{}p{15cm}@{}}
\toprule
You are a helpful assistant that is helping web developers to design the color of their websites. With the provided color palette and the code (HTML and CSS), re-assign the best colors to the UI elements in the code. To ensure accessible visual contrast, apply colors only in the intended pairs or layering orders described below. Consider the elements' overlay relationship and apply colors appropriately. Do not remove any existing CSS class. Do not change other CSS attributes than color-related attributes (e.g., width, padding).\\
\textbf{Definitions}\\
\# Primary: Use primary roles for the most prominent components across the UI, such as the FAB, high-emphasis buttons, and active states.\\
\# Colors: \\
\# [1] primary: High-emphasis fills, texts, and icons against surface\\
\# [2] on\_primary: Text and icons against primary\\
\# [3] primary\_container: Standout fill color against surface, for key components like FAB\\
\# [4] on\_primary\_container: Text and icons against primary container\\

\# Secondary: Use secondary roles for less prominent components in the UI such as filter chips.\\
\# Colors: \\
\# [1] secondary: Less prominent fills, text, and icons against surface\\
\# [2] on\_secondary: Text and icons against secondary\\
\# [3] secondary\_container: Less prominent fill color against surface, for recessive components like tonal buttons\\
\# [4] on\_secondary\_container: On secondary container – Text and icons against secondary\\

\# Tertiary: Use tertiary roles for contrasting accents that balance primary and secondary colors or bring heightened attention to an element such as an input field. The tertiary color roles can be applied at the designer's discretion. They're intended to support broader color expression.\\
\# Colors: \\
\# [1] tertiary: Complementary fills, text, and icons against surface\\
\# [2] on\_tertiary: Text and icons against tertiary\\
\# [3] tertiary\_container: Complementary container color against surface, for components like input fields\\
\# [4] on\_tertiary\_container: Text and icons against tertiary container\\

\# Error: Use error roles to communicate error states, such as an incorrect password entered into a text field\\
\# Colors:\\
\# [1] error: Attention-grabbing color against surface for fills, icons, and text, indicating urgency\\
\# [2] on\_error: Text and icons against error\\
\# [3] error\_container: Attention-grabbing fill color against surface\\
\# [4] on\_error\_container: Text and icons against error container\\

\# Surface: Use surface roles for more neutral backgrounds, and container colors for components like cards, sheets, and dialogs. The most common combination of surface roles uses surface for a background area and surface container for a navigation area. For example, the body area will use the surface color and the navigation area will use the surface container color on both mobile and tablet.\\
\# Colors: \\
\# [1] surface: Default color for backgrounds, \\
\# [2] on\_surface: Text and icons against any surface color\\

\# Outline: Subtle color used for boundaries\\
\# Colors: \\
\# [1]: outline: Important boundaries, such as a text field outline\\\\

\textbf{Input Color Palette:} // Palette generated using Materials Color Utilities //\\
primary: \#506600, on\_primary: \#ffffff\\
...\\

\textbf{Input HTML:} \\
...\\

\textbf{Input CSS:} \\
...\\
\textbf{Output CSS:} \\
\bottomrule
\end{tabular}
}
\caption{A full example of prompt used to provide color palette suggestions.}
\end{table*}

\clearpage
\subsection{Website Dataet for Capturing Design Trends}~\label{apndx:website_dataset}
To capture the trend of different types of websites, we constructed a dataset of 100 websites of 5 different categories: Blog, Tutorial, Personal Website, Organization Website, and News/Magazine (full list of websites in Supplementary Materials). While website designs may vary greatly depending on the content and purpose~\cite{kumar2013webzeitgeist}, we focus on improving the design of these types, which, as highlighted in our formative study, are predominantly created by BLV developers. One researcher compiled a list of websites by referencing online website listings for each type (\textit{e.g.,} Searching ``Top 10 design for blog websites'').  From the first 100 search results, we randomly selected 20 for each category. During the selection process, we filtered out websites with ad popups that obstruct full access to the content and duplicate websites from the same source (\textit{e.g.,} multiple entries from NY Times), to ensure a wide range of unique design examples. Websites that haven't been updated in over five years were excluded from the dataset to capture current web design trends. 
\\\\\\

\section{USER EVALAUATION MATERIALS}

\begin{figure}[htbp!]
  \centering
  \includegraphics[width=5in]{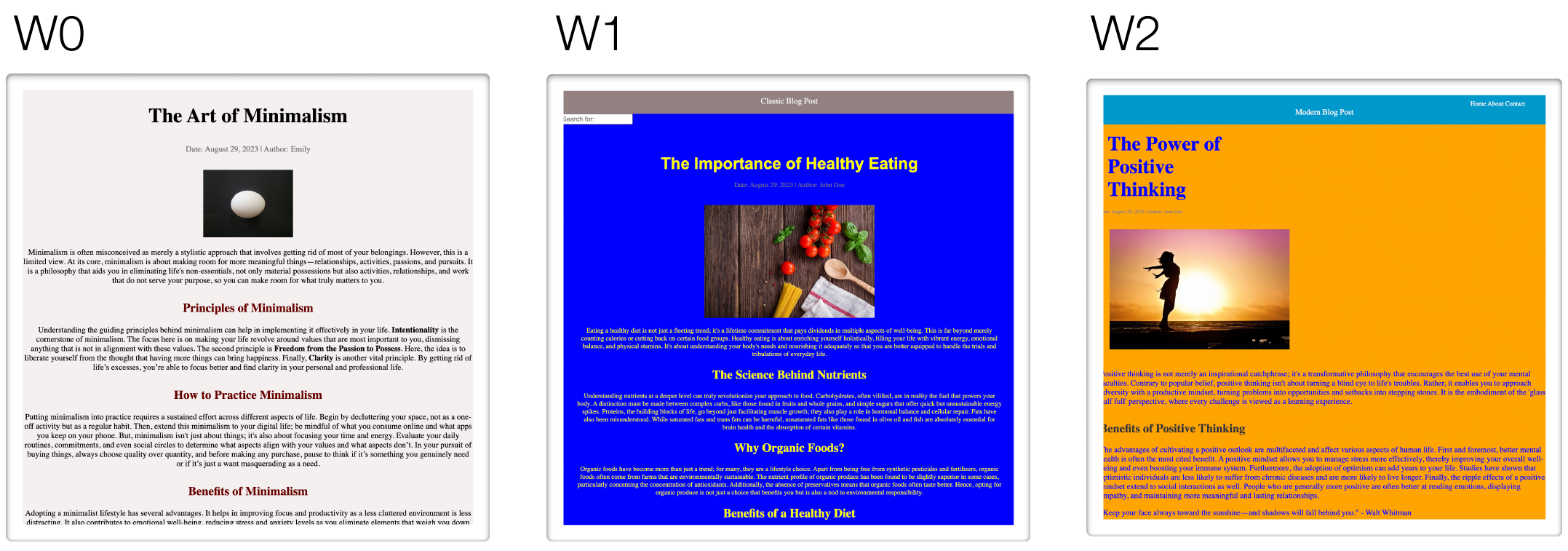}
  \caption{Website materials used in the evaluation study. W0 was used for the tutorial session and W1 and W2 were used for the editing task. See Supplementary Materials for the full code of W0-W2. }
  \label{fig:materials}
\end{figure}

\end{document}